%% file: main.tex
\documentclass[5p]{elsarticle}
\usepackage{filecontents}

\usepackage{setspace}
\usepackage{balance}

\usepackage{listings}
\usepackage{syntax}
\usepackage{booktabs}
\usepackage{color, colortbl}
\usepackage{xcolor}
\definecolor{RowHighlight}{rgb}{0.88,1,1}
\definecolor{bananamania}{rgb}{0.98, 0.91, 0.71}
\usepackage{hyperref}
\usepackage{amsthm}
\usepackage{multicol}
\usepackage{graphicx}
\usepackage{verbatim}
\usepackage{subcaption}

\usepackage{ragged2e}
\usepackage{blindtext}
\usepackage{todonotes}
\usepackage{amsmath}
\usepackage{colornames}
\usepackage{soul,color}

\sethlcolor{bananamania}

\def\tsc#1{\csdef{#1}{\textsc{\lowercase{#1}}\xspace}}
\tsc{WGM}
\tsc{QE}
\tsc{EP}
\tsc{PMS}
\tsc{BEC}
\tsc{DE}
\newcommand{\kw}[1]{\textcolor{blue}{\textsf{#1}}}
\newcommand{\cspkw}[1]{\textbf{\color{black} #1}}

\definecolor{codeBack}{RGB}{255, 253,232}
\definecolor{eclipse}{RGB}{127,0,85}

\lstdefinelanguage{SLEEC}{
	morekeywords={constant, measure, event, true, false, def_start, def_end, rule_start, rule_end, scale, real, int, bool, parameter, min, max, when, then, within, withProb, otherwise, also, and, while, or, not, unless, seconds, minutes},
	sensitive=false, % keywords are not case-sensitive
	morecomment=[l]{//}, % l is for line comment
	morecomment=[s]{/*}{*/}, % s is for start and end delimiter
	morestring=[b]", % defines that strings are enclosed in double quotes
 	basicstyle=\fontsize{7.6}{7.6}\selectfont\ttfamily,
	keywordstyle=\color{blue},
	tabsize=2,
	commentstyle=\color{gray}
} %

\lstdefinelanguage{xtend}{
morekeywords={val, private, def, if, else, then, null},
sensitive=false,
basicstyle=\fontsize{7.6}{7.6}\selectfont\ttfamily,
keywordstyle=\color{eclipse},
	tabsize=2
}

\lstdefinelanguage{BNF}{
    keywords={},
    otherkeywords={% Operators
        >, <, |, `, '
    },
	sensitive=false, % keywords are not case-sensitive
	morecomment=[l]{//}, % l is for line comment
	morecomment=[s]{/*}{*/}, % s is for start and end delimiter
	morestring=[b]", % defines that strings are enclosed in double quotes
	basicstyle=\fontsize{8}{8}\selectfont\ttfamily,
	keywordstyle=\color{blue},
	keywordstyle=[2]\color{blue},% for example
	tabsize=2,
	commentstyle=\color{gray},
	xleftmargin=0.2cm,
	xrightmargin=0.2cm
} %

\newcommand{\nterm}[1]{\textbf{\textsf{#1}}}

\newcommand{\acronym}{SLEEC~}
\theoremstyle{definition}
\newtheorem{example}{Example}[section]
\newtheorem{definition}{Definition}[section]

\usepackage{amssymb}
\usepackage[zed,color]{csp}

\definecolor{ZedColor}{rgb}{0,0,0}
\definecolor{MetaColor}{rgb}{1,0,0}
\definecolor{GreyColor}{rgb}{0.5,0.5,0.5}

\newcommand{\cspl}[1]{#1}%{\color{ZedColor}#1}}
\newcommand{\cseq}{\mathrel{;}\mkern-5.5mu}
\newcommand{\RC}[1]{\textsf{#1}}
\newcommand{\meta}[1]{\mathsf{#1}}%\color{GreyColor}{#1}}}

\newcounter{tRule}%[section]

\newcommand{\lsem}{\meta{\lbrack\!\lbrack}}
\newcommand{\rsem}{\rbrack\!\rbrack}
\newcommand{\rsemR}{\meta{\rsem_R}}
\newcommand{\rsemS}{\meta{\rsem_S}}
\newcommand{\rsemDS}{\meta{\rsem_{DS}}}
\newcommand{\rsemD}{\meta{\rsem_D}}
\newcommand{\rsemT}{\meta{\rsem_T}}
\newcommand{\rsemRS}{\meta{\rsem_{RS}}}
\newcommand{\rsemTG}{\meta{\rsem_{TG}}}
\newcommand{\rsemME}{\meta{\rsem_{ME}}}
\newcommand{\rsemRDS}{\meta{\rsem_{RDS}}}
\newcommand{\rsemRP}{\meta{\rsem_{C}}}
\newcommand{\rsemCDS}{\meta{\rsem_{CDS}}}
\newcommand{\rsemEDS}{\meta{\rsem_{EDS}}}
\newcommand{\rsemLRDS}{\meta{\rsem_{LRDS}}}

\newcommand{\rsemCP}{\meta{\rsem_{CP}}}

\newcommand{\startBy}{\blacktriangleleft}
\newcommand{\after}{\mathrel{\mathrm{after}}}
\hypersetup{pdfauthor={Name}}

\begin{document}
%%%%%%%%%%%%%%%%%%%%%%%%%%%%%%%%%%%%%%%%%%%%%%%%%%%%%%%%%%%%
%%%%%%%%%%%%%%%%%%%%%%%%%%%%%%%%%%%%%%%%%%%%%%%%%%%%%%%%%%%%
\begin{frontmatter}

\title{Specification, Validation and Verification of
\\
Social, Legal, Ethical, Empathetic and Cultural
Requirements 
\\
for Autonomous Agents}

\author{Sinem Getir Yaman$^1$, Ana Cavalcanti$^1$, Radu Calinescu$^1$, Colin Paterson$^1$, Pedro Ribeiro$^1$, and Beverley Townsend$^2$}

\address{1. Department of Computer Science, University of York, UK\\
2. York Law School, University of York, UK}

\begin{singlespace}
\begin{abstract}
Autonomous agents are increasingly being proposed for use in healthcare, assistive care, education, and other applications governed by complex human-centric norms. To ensure compliance with these norms, the rules they induce need to be unambiguously defined, checked for consistency, and used to verify the agent. In this paper, we introduce a framework for formal specification, validation and verification of social, legal, ethical, empathetic and cultural (SLEEC) rules for autonomous agents. Our framework comprises:~(i)~a language for specifying SLEEC rules and rule \emph{defeaters} (that is, circumstances in which a rule does not apply or an alternative form of the rule is required); (ii)~a formal semantics (defined in the process algebra tock-CSP) for the language; and (iii)~methods for detecting conflicts and redundancy within a set of rules, and for verifying the compliance of an autonomous agent with such rules. We show the applicability of our framework for two autonomous agents from different domains:~a firefighter UAV, and an assistive-dressing robot.
\end{abstract}

\end{singlespace}
\end{frontmatter}

%\linenumbers

%%%%%%%%%%%%%%%%%%%%%%%%%%%%%%%%%%%%%%%%%%%%%%%%%%%%%%%%%%%%
%%%%%%%%%%%%%%%%%%%%%%%%%%%%%%%%%%%%%%%%%%%%%%%%%%%%%%%%%%%%
\section{Introduction}
%%%%%%%%%%%%%%%%%%%%%%%%%%%%%%%%%%%%%%%%%%%%%%%%%%%%%%%%%%%%
%%%%%%%%%%%%%%%%%%%%%%%%%%%%%%%%%%%%%%%%%%%%%%%%%%%%%%%%%%%%

\input{introduction}

%%%%%%%%%%%%%%%%%%%%%%%%%%%%%%%%%%%%%%%%%%%%%%%%%%%%%%%%%%%%
%%%%%%%%%%%%%%%%%%%%%%%%%%%%%%%%%%%%%%%%%%%%%%%%%%%%%%%%%%%%
\section{Running example}
\label{sec:runexample}
%%%%%%%%%%%%%%%%%%%%%%%%%%%%%%%%%%%%%%%%%%%%%%%%%%%%%%%%%%%%
%%%%%%%%%%%%%%%%%%%%%%%%%%%%%%%%%%%%%%%%%%%%%%%%%%%%%%%%%%%%

\input{example}

%%%%%%%%%%%%%%%%%%%%%%%%%%%%%%%%%%%%%%%%%%%%%%%%%%%%%%%%%%%%
%%%%%%%%%%%%%%%%%%%%%%%%%%%%%%%%%%%%%%%%%%%%%%%%%%%%%%%%%%%%
\section{The \acronym Language}
\label{sec:language}
%%%%%%%%%%%%%%%%%%%%%%%%%%%%%%%%%%%%%%%%%%%%%%%%%%%%%%%%%%%%
%%%%%%%%%%%%%%%%%%%%%%%%%%%%%%%%%%%%%%%%%%%%%%%%%%%%%%%%%%%%

\input{language}

%%%%%%%%%%%%%%%%%%%%%%%%%%%%%%%%%%%%%%%%%%%%%%%%%%%%%%%%%%%%
%%%%%%%%%%%%%%%%%%%%%%%%%%%%%%%%%%%%%%%%%%%%%%%%%%%%%%%%%%%%
\section{SLEEC Semantics} 
\label{sec:semantics}
%%%%%%%%%%%%%%%%%%%%%%%%%%%%%%%%%%%%%%%%%%%%%%%%%%%%%%%%%%%%
%%%%%%%%%%%%%%%%%%%%%%%%%%%%%%%%%%%%%%%%%%%%%%%%%%%%%%%%%%%%

\input{semantics.tex}

%%%%%%%%%%%%%%%%%%%%%%%%%%%%%%%%%%%%%%%%%%%%%%%%%%%%%%%%%%%%
%%%%%%%%%%%%%%%%%%%%%%%%%%%%%%%%%%%%%%%%%%%%%%%%%%%%%%%%%%%%
\section{Validation and Verification}
\label{sec:validation}
%%%%%%%%%%%%%%%%%%%%%%%%%%%%%%%%%%%%%%%%%%%%%%%%%%%%%%%%%%%%
%%%%%%%%%%%%%%%%%%%%%%%%%%%%%%%%%%%%%%%%%%%%%%%%%%%%%%%%%%%%

\input{vandv.tex}

%%%%%%%%%%%%%%%%%%%%%%%%%%%%%%%%%%%%%%%%%%%%%%%%%%%%%%%%%%%%
%%%%%%%%%%%%%%%%%%%%%%%%%%%%%%%%%%%%%%%%%%%%%%%%%%%%%%%%%%%%
% alcc: commented here and included in evaluation.tex to assist with the management of figure positioning. 
%\section{Evaluation: Tool Support and Case Studies}
%\label{sec:evaluation}
%%%%%%%%%%%%%%%%%%%%%%%%%%%%%%%%%%%%%%%%%%%%%%%%%%%%%%%%%%%%
%%%%%%%%%%%%%%%%%%%%%%%%%%%%%%%%%%%%%%%%%%%%%%%%%%%%%%%%%%%%

\input{evaluation.tex}

%%%%%%%%%%%%%%%%%%%%%%%%%%%%%%%%%%%%%%%%%%%%%%%%%%%%%%%%%%%%
%%%%%%%%%%%%%%%%%%%%%%%%%%%%%%%%%%%%%%%%%%%%%%%%%%%%%%%%%%%%
\section{Related work}
\label{sec:relatedwork}
%%%%%%%%%%%%%%%%%%%%%%%%%%%%%%%%%%%%%%%%%%%%%%%%%%%%%%%%%%%%
%%%%%%%%%%%%%%%%%%%%%%%%%%%%%%%%%%%%%%%%%%%%%%%%%%%%%%%%%%%%

\noindent
A commonality in normative themes found in many recently developed artificial intelligence ethics and guidance instruments inform a ‘normative core’ of a principled approach to development, deployment, and adoption~\cite{jobin2019global}~\cite{UNESCO-Ethics} ~\cite{OECD-Recommendation} of agents whose behaviour relies on use of artificial intelligence techniques. Significant work has been done in the development of autonomous systems from the perspective of normative ideas ~\cite{fjeld2020principled,dennis2015towards} including work around transparency ~\cite{IEEETransparency}, explainability, and accountability~\cite{Inverardi22}. Another research perspective develops a data-driven personalised tool, based on the moral choices of the user~\cite{AlfieriIMP22}. 
Our SLECC framework, however, is concerned with the problem of the operationalisation of such norms~\cite{TownsendPANCCHT22}~\cite{SolankiGH23} and defines a formalisation and an automated process for validating and verifying rules that capture these norms.

There exists significant research on the development and verification of autonomous systems~\cite{LuckcuckMattetalSURVEY}\cite{Normann2014HW}\cite{menghi2019specification}. Most of the approaches verify the autonomous agents using formal verification methods, such as model checking and theorem proving, by introducing new formalisms, but---com\-ple\-men\-tary to our SLEEC framework---they focus on the safety requirements of the agents. 

Concerns with verification regarding some level of ethical constraints and legal aspects~\cite{Bhuiyan2020AMF} have been recently studied~\cite{Bremner2019DFW, Dennis2016FSW} and  investigated from a verification perspective~\cite{dennis2015towards}, although not from the perspective of operationalisation of these requirements. Robots are formally verified in~\cite{dennis2015towards} with an action selection of the robot controller which evaluates the outcomes of actions using simulation and prediction, and makes selection using a safety/ethical logic~\cite{Bremner2019DFW}. Bremer et al.~\cite{Bremner2019DFW} present a technique for verification of transparency and ethical concerns using the belief–desire–intention model and a simulation module to obtain ethical rules. This line of work is complementary to our SLEEC framework, as our focus is not on the identification of rules. Furthermore, unlike our SLEEC framework, these approaches do not provide a notation dedicated to the encoding of SLEEC-related concerns as requirements.

%%%%%%%%%%%%%%%%%%%%%%%%%%%%%%%%%%%%%%%%%%%%%%%%%%%%%%%%%%%%
%%%%%%%%%%%%%%%%%%%%%%%%%%%%%%%%%%%%%%%%%%%%%%%%%%%%%%%%%%%%
\section{Conclusion}

We introduced a tool-supported framework for the end-to-end specification, consistency validation and verification of social, legal, ethical, empathetic and cultural requirements for autonomous agents. The framework supports the specification of these requirements as SLEEC rules formalised in a domain-specific language grounded in defeasible logic~\cite{horty2012reasons,zalta2005stanford}, and their translation into tock-CSP for redundancy and conflict checking, and for verifying autonomous agent compliance with SLEEC requirements. By enabling the operationalisation of SLEEC requirements for autonomous agents, our framework complements the significant international efforts to define ethical principles for AI and autonomous systems, e.g.~\cite{UNESCO-Ethics,OECD-Recommendation,IEEE-EAD}, and our own recent work to elicit SLEEC requirements for autonomous agents by starting from relevant normative principles and stakeholder needs~\cite{TownsendPANCCHT22}.

In future work, we will explore several opportunities for extending the applicability and usability of our SLEEC framework. First, we plan to augment the SLEEC language with probabilistic constructs, and thus to provide support for modelling the uncertainty in the environment and decisions of autonomous agents. Second, we will improve the integration of the tools used by the framework, e.g., through automating the invocation of the FDR model checker used to verify autonomous agent compliance with SLEEC requirements. Third, we intend to augment our tool support with a module that converts the counterexample traces produced by the model checker into error messages that are easier to understand for framework users who do not have formal methods expertise. Finally, we plan to continue to evaluate the SLEEC framework in additional case studies from different application domains, and with a larger number of SLEEC experts (lawyers, ethicists, domain experts, regulators, sociologists) in order to identify and fix any remaining applicability and usability issues, and to assess the scalability and generalisability of the framework further.

In the longer term, we will consider also extending the framework with two additional methods. The former method is needed for the runtime verification of autonomous-agent decisions. Many autonomous systems learn, adapt and evolve in operation, e.g., in response to changes in their environment, and therefore cannot be fully verified at development time. The latter method will further support this evolution by focusing on the online synthesis of SLEEC-compliant adaptation plans for autonomous agents.

\label{sec:discussion}
%%%%%%%%%%%%%%%%%%%%%%%%%%%%%%%%%%%%%%%%%%%%%%%%%%%%%%%%%%%%
%%%%%%%%%%%%%%%%%%%%%%%%%%%%%%%%%%%%%%%%%%%%%%%%%%%%%%%%%%%%

%%%%%%%%%%%%%%%%%%%%%%%%%%%%%%%%%%%%%%%%%%%%%%%%%%%%%%%%%%%%
%%%%%%%%%%%%%%%%%%%%%%%%%%%%%%%%%%%%%%%%%%%%%%%%%%%%%%%%%%%%
\section*{Acknowledgements} 
The work described in this paper was funded by the 

EPSRC project EP/V026747/1 `UKRI Trustworthy Autonomous Systems Node in Resilience’. The work of Ana Cavalcanti is funded by the Royal Academy of Engineering grant CiET1718/45 and `UKRI TAS
Verifiability Node EP/V026801/1'. The work of Pedro Ribeiro is funded by EPSRC RoboTest EP/R025479/1. 
%\balance

\bibliographystyle{elsarticle-num}

\end{document}

%% file: introduction.tex
There is huge push to develop and use autonomous agents (software and cyber-physical systems) in high-stakes applications from health and social care, transportation, education, and other domains.
Along functional and non-functional requirements such as dependability, performance and utility, a new class of non-functional requirements related to social, legal, ethical, empathetic, and cultural (SLEEC) concerns~\cite{TownsendPANCCHT22} has become increasingly important and challenging for these applications ~\cite{wing2021trustworthy,moor2006nature,Inverardi22}. %
Despite that recognised importance, there is currently very little support for the elicitation, specification, validation, and verification of SLEEC requirements. Existing research in the area is promising, but only covers specific aspects of the problem. For example, there are results on the  study~\cite{Bremner2019DFW, Dennis2016FSW} and verification~\cite{dennis2015towards} of ethical concerns of autonomous agents, modelling of legal requirements for software systems~\cite{boltz2022model}, and development of personalised ethical assistant tools based on the moral choices of the user~\cite{AlfieriIMP22}. 

In this paper, we build on this early research to provide support for the development of autonomous agents that need to perform tasks that raise SLEEC concerns~\cite{TownsendPANCCHT22,Floridi2018}. %\textcolor{green}{added:covering both soft and hard ethics perspective}~\cite{Floridi2018}. 
To that end, we introduce a tool-supported SLEEC requirement specification and verification framework that includes a language for defining these concerns as \emph{SLEEC rules} that complement the functional and other non-func\-tional requirements of an autonomous agent. Our language supports the use of defeasible logic~\cite{horty2012reasons,zalta2005stanford} to 
allow both the definition of SLEEC constraints and the specification of conditions under which these constraints do not apply or may need to be replaced with alternative constraints. Such conditions are expressed in terms of additional information coming from the environment or the agent components, and are specified within SLEEC rules as \emph{defeaters}. 
Given a set of SLEEC rules for an autonomous agent, our framework automates: (i)~their formalisation in tock-CSP~\cite{BaxterRC22}, a version of the communicating sequential processes (CSP) algebra~\cite{Hoare78CSP} that can describe discrete-time properties;
(ii)~the validation of their consistency, to ensure that the rules are not conflicting, and to identify redundant rules; and (ii)~the verification of an agent's compliance with the validated rules.  

We have evaluated our framework with two case studies:~a firefighter uncrewed aerial vehicle (UAV) and an assistive robot application from the healthcare domain. Their rules have been identified with the help of lawyers and ethicists. Our models representing the agent behaviour are developed using a domain-specific language for robotics, RoboChart~\cite{Robochart}. This is a diagrammatic notation that can be used to model control software using state machines, time primitives to capture budgets and deadlines, and a simple component model.  Since there is support to generate tock-CSP models of RoboChart diagrams automatically, we can use these models to formally verify designs of the autonomous agents' software against SLEEC rules. 

The main contributions of our paper include: 
\begin{enumerate}[(1)]
\item A domain-specific language supporting the specification of SLEEC rules for autonomous agents.
\item The definition of a formal semantics for this language in tock-CSP, catering for the definition of time budgets, deadlines, and timeouts in the rules.
\item A method for the formal validation of SLEEC specifications, to detect conflicting and redundant rules.
\item A method for formally verifying the compliance of a tock-CSP-encoded agent specification or design with respect to a set of valid SLEEC rules.
\item End-to-end tool support for SLEEC requirements specification, consistency validation and verification, using a combination of software components developed by our project and the FDR model checker~\cite{FDR}.
\end{enumerate}

\noindent%
The remainder of the paper is structured as follows. Section \ref{sec:runexample} introduces the firefighter UAV, which we use as a running example in later sections. Section~\ref{sec:language} presents the SLEEC language, and 
Section~\ref{sec:semantics} defines its formal semantics. Section~\ref{sec:validation} describes our approach to conflict and redundancy checking, and our verification process for SLEEC specifications. Section~\ref{sec:evaluation} details our evaluation, describing the tool support provided, and the two case studies. Finally, Section~\ref{sec:relatedwork} covers related work, and Section~\ref{sec:discussion} concludes the paper with a brief summary and a discussion of directions for future work.

%% file: example.tex
To illustrate the concepts, notation, methods, and application of our SLEEC framework, we use as a running example a firefighting UAV inspired by recent research on the use of drones to help tackle wildfires and urban fires~\cite{alon2021drones,cervantes2018conceptual,innocente2019self}. We consider that this UAV is tasked with: (i)~using a thermal camera to detect a potential fire at a warehouse; (ii)~determining the precise location of the fire~(with its depth camera) to report to a human teleoperator; and (iii)~using an onboard water spraying system to control the fire until the arrival of the fire brigade. 

In addition to these functional goals, we suppose that the firefighter UAV from our running example needs to consider SLEEC concerns arising from its interactions with human firefighters, bystanders and teleoperators. For example, we assume that the UAV has an alarm which sounds when the battery is running low. However, there are social concerns about sounding a loud alarm too close to a human.  
As another example, we consider that reporting a (potential) fire involves sending video footage of the surveyed building to teleoperators. If, however, bystanders are present in the vicinity of the building, including them in this footage can raise legal and/or ethical privacy concerns.  We explain the UAV capabilities and the associated SLEEC concerns in detail as we introduce our SLEEC notation and its semantics in the next sections.

%% file: language.tex
%%%%%%%%%%%%%%%%%%%%%%%%%%%%%%%%%%%%%%%%%%%%%%%%%%%%%%%%%%%%
%%%%%%%%%%%%%%%%%%%%%%%%%%%%%%%%%%%%%%%%%%%%%%%%%%%%%%%%%%%%

Our framework supports the definition of SLEEC rules for an autonomous agent by using a domain-specific language whose syntax is defined in Figure~\ref{lst:BNF}. The set of SLEEC rules for an agent is provided in this language as a \textbf{specification} comprising two blocks. The first block~(an element of the syntactic category~\textbf{defBlock}) provides \textbf{definitions} for the functional capabilities and parameters of the agent. The second block~(\textbf{ruleBlock}) defines the actual SLEEC \textbf{rules} in terms of those capabilities and parameters. These blocks are described next.

\begin{figure*}\centering
\setlength{\tabcolsep}{0.7\tabcolsep}

\begin{tabular}{l@{\ } l@{\ } p{123mm}}
\toprule
\nterm{specification} & $::=$ & \nterm{defBlock} \nterm{ruleBlock}\\ \midrule
\nterm{defBlock} & $::=$ & \kw{def_start} \nterm{definitions} \kw{def_end}\\
\nterm{definitions} & $::=$ & \nterm{definition} $|$ \nterm{definition} \nterm{definitions}\\
\nterm{definition} & $::=$ & \kw{event} \nterm{eventID} $|$ 
                 \kw{measure} \nterm{measureID} : \nterm{type} $|$
                 \kw{constant} \nterm{constID} [= \nterm{value}] \\
\nterm{type} & $::=$ & \kw{boolean} $|$ \kw{numeric} $|$ \kw{scale}(\nterm{scaleParams})\\
\nterm{scaleParams} & $::=$ & \nterm{literal} $|$ \nterm{literal}, \nterm{scaleParams}\\ \midrule
\nterm{ruleBlock} & $::=$ & \kw{rule_start} \nterm{rules} \kw{rule_end}\\
\nterm{rules} & $::=$ & \nterm{rule} $|$ \nterm{rule} \nterm{rules}\\
\nterm{rule} & $::=$ & \nterm{ruleID} \kw{when} \nterm{trigger} \kw{then} \nterm{response} \\
\nterm{trigger} & $::=$ & \nterm{eventID} $|$ \nterm{eventID} \kw{and} \nterm{mBoolExpr}\\
\nterm{mBoolExpr} & $::=$ & \nterm{measureID} $|$ \kw{not} \nterm{mBoolExpr} 
                $|$ (\nterm{mBoolExpr})
                $|$ \nterm{mBoolExpr} \nterm{relOp} \nterm{value}
                $|$ \\
                & & \nterm{mBoolExpr} \nterm{boolOp} \nterm{boolValue}\\
\nterm{response} & $::=$ & \nterm{constraint} [\nterm{defeaters}] $|$ \{ \nterm{constraint} [\nterm{defeaters}] \} \\
\nterm{constraint} & $::=$ & \nterm{eventID} [\kw{within} \nterm{value} \nterm{timeUnit} [\kw{otherwise} \nterm{response}]]\\
& & $|$ \kw{not} \nterm{eventID} \kw{within} \nterm{value} \nterm{timeUnit} \\
\nterm{defeaters} & $::=$ & \nterm{defeater} $|$ \nterm{defeater} \nterm{defeaters} \\
\nterm{defeater} & $::=$ & \kw{unless} \nterm{mBoolExpr} [\kw{then} \nterm{response}] \\
\nterm{relOp} & $::=$ & $<$ $|$ $>$ $|$ $<>$ $|$ $<=$ $|$ $>=$ $|$ $=$\\
\nterm{boolOp} & $::=$ & \kw{and} $|$ \kw{or}\\
\bottomrule
\end{tabular}
\caption{BNF syntax of the \acronym language \label{lst:BNF}
}
\end{figure*}
%%%%%%%%%%%%%%%%%%%%%%%%%%%%%%%%%%%%%%%%%%%%%%%%%%%%%%%%%%%%
%%%%%%%%%%%%%%%%%%%%%%%%%%%%%%%%%%%%%%%%%%%%%%%%%%%%%%%%%%%%
\subsection{The definitions block}
%%%%%%%%%%%%%%%%%%%%%%%%%%%%%%%%%%%%%%%%%%%%%%%%%%%%%%%%%%%%
%%%%%%%%%%%%%%%%%%%%%%%%%%%%%%%%%%%%%%%%%%%%%%%%%%%%%%%%%%%%

The \textbf{definitions} block~(delimited by the keyword pair \kw{def\_start}\ldots\kw{def\_end}) comprises declarations of \kw{event}s and \kw{measure}s that represent capabilities of the agent, and \kw{constant}s that represent parameters of the agent. Events and measures correspond to interactions between the agent and the environment, including any humans, to reflect aspects of the environment that are perceptible or affected by the agent. Measures differ from events in that they carry values, communicated to the agent on demand. A measure corresponds to a query that is always responsive.  An event is an atomic interaction~(input or output) that happens sporadically.  

A \kw{constant} represents a value for some parameter of the system configuration; its specific value may or may not be defined. If a value  is not defined, the constant represents, for instance, a parameter that is defined at deployment time to reflect the hardware or environment in which the system is deployed, or the preferences of its user. 

\begin{example}
An example of a definition block for our firefighter UAV is shown in Listing~\ref{lst:defBlock}. \textsf{BatteryCritical} is an event that occurs when the battery is very low. This is an abstraction for a battery sensor that provides input to the UAV regarding its own hardware. \textsf{CameraStart} represents an interaction with a teleoperator, who can turn on the camera and start recording. \textsf{SoundAlarm} is associated with a loudspeaker that the UAV can use to sound an alarm. Finally, \textsf{GoHome} represents a navigation capability of the UAV, provided by its motors and the embedded software for using these motors to return the UAV to a home location. 

In addition, the SLEEC definition block from Listing~\ref{lst:defBlock} defines three measures. The first, \textsf{personNearby}, communicates a boolean to indicate whether, using its cameras and associated vision software, the firefighter UAV has detected the presence of a person. Whenever that information is needed, the agent can use the \textsf{personNearby} measure to obtain it. We declare also two measures for the \textsf{temperature} of the air, and the \textsf{windSpeed} level. 
Finally, the constant \textsf{ALARM\_DEADLINE} records a ``time budget'' for the alarm to sound. We do not give its value in the specification, as we assume it is dependent on the actual deployment of the UAV.

\hfill $\Box$
\end{example}

\begin{lstlisting}[language = SLEEC, caption={Definition block for our firefighter robot.}, label={lst:defBlock},captionpos=b, float=tp, frame = single]
def_start
  event BatteryCritical
  event CameraStart
  event SoundAlarm
  event GoHome
  measure personNearby: boolean 
  measure temperature: numeric
  measure windSpeed: scale(light,moderate,strong)
  constant ALARM_DEADLINE 
def_end
\end{lstlisting}

\noindent
In summary, an event can be issued by an agent; \textsf{GoHome} is an example. Alternatively, an event can be a request issued by a user, such as \textsf{CameraStart}, or an input from another system component, such as \textsf{BatteryCritical}. Measures, on the other hand, provide to the agent information about the state of the system. Some measures may be known with a high degree of certainty from sensors, such as a temperature sensor or a heart-rate monitor. Others may be inferred from indirect measures or indeed the fusion of multiple sensors. For instance, a user's level of distress may be inferred from heart-rate monitors, images of the user's facial expression, and their tone of voice. 

Events, measures, and constants have a unique identifier~(\textbf{eventID}, \textbf{measureID}, and \textbf{constID}). By convention, we use identifiers starting with a capital letter for events, and with a lowercase letter for measures. For constants, we use identifiers all in capitals. A measure declaration also defines the \textbf{type} of the values it communicates.  The supported types are \kw{boolean}, \kw{numeric}, and ordinal \kw{scale}s, which introduce some literals and define an order among them. In our example, we have \kw{scale}\textsf{(light,moderate, strong)} with the implicit order \textsf{light $<$ moderate $<$ strong}. 

%%%%%%%%%%%%%%%%%%%%%%%%%%%%%%%%%%%%%%%%%%%%%%%%%%%%%%%%%%%%
%%%%%%%%%%%%%%%%%%%%%%%%%%%%%%%%%%%%%%%%%%%%%%%%%%%%%%%%%%%%
\subsection{The rules block
}
%%%%%%%%%%%%%%%%%%%%%%%%%%%%%%%%%%%%%%%%%%%%%%%%%%%%%%%%%%%%
%%%%%%%%%%%%%%%%%%%%%%%%%%%%%%%%%%%%%%%%%%%%%%%%%%%%%%%%%%%%

SLEEC \textbf{rules} are defined in a \textbf{ruleBlock}~(delimited by the keywords \kw{rule\_start}\ldots\kw{rule\_end}). A rule has an identifier~(\textbf{ruleID}), a \textbf{trigger}, and a \textbf{response}. A \textbf{trigger} is an event and, optionally, a condition (i.e., a Boolean expression) over measures~(\textbf{mBoolExpr}); \kw{when} the event in the \textbf{trigger} occurs and the condition, if any, is satisfied, \kw{then} the rule specifies the required \kw{response} which defines a \textbf{constraint} indicating the event(s) that must or must not occur.  The Boolean expression over measures can include conjunctions~(\kw{and}), disjunctions~(\kw{or}), and equalities and inequalities~(\textbf{relOp}) over \kw{numeric} and \kw{scale} measures and relevant values.

\begin{example}
In Listing~\ref{lst:trigger}, \textsf{Rule1} is concerned with the privacy of persons near the firefighter UAV, when its camera starts recording. The trigger of \textsf{Rule1} has the event \textsf{CameraStart} and a condition requiring the value of the measure \textsf{personNearby} to be true.  The response, in this case, consists of a constraint that requires \textsf{SoundAlarm} to occur, to warn the person that recording is underway. \hfill $\Box$
\end{example}

\begin{lstlisting}[language = SLEEC, label={lst:trigger}, caption={Sample SLEEC rules for a firefighter UAV},captionpos=b, float=tp, frame = single]
rule_start
  Rule1 when CameraStart and personNearby 
        then SoundAlarm
  Rule2 when CameraStart and personNearby 
        then SoundAlarm within 2 seconds
  Rule3 when SoundAlarm
        then not GoHome within 5 minutes 
  Rule4 when CameraStart then SoundAlarm 
        unless personNearby then GoHome
        unless temperature > 35 
rule_end
\end{lstlisting}

\noindent
A distinctive feature of our SLEEC language is that it can specify time constraints:~time budgets for responses and required alternative responses in the case of a timeout. A time budget is specified using the \kw{within} construct. The \textbf{timeUnit} is provided based on the context under consideration. 

\begin{example}
  In Listing~\ref{lst:trigger}, \textsf{Rule2} is a more specific variant of \textsf{Rule1}.  It has the same trigger as \textsf{Rule1}, but gives a time budget for the response:~it must happen \kw{within} \textsf{2} \kw{seconds}. After all, if the person is not warned early enough, the recording might already have violated their privacy by the time the alarm sounds. \hfill $\Box$  
\end{example}

\noindent%
For situations where the response event may not happen within its budget, i.e., when there is a timeout, the \kw{otherwise} construct can be used to define an alternative response.

\begin{example}
  Revisiting \textsf{Rule2} from Listing~\ref{lst:trigger}, we may realise that achieving \textsf{SoundAlarm} within 2~seconds is not guaranteed. The loudspeakers may be broken, or another SLEEC rule may specify that sounding the alarm is not a socially or ethically acceptable course of action, e.g., because the person is too close to the UAV.  In this case, an alternative can be provided as shown in \textsf{Rule2\_a} from Listing~\ref{lst:trigger2}. Here, the UAV is required to return to base~(\textsf{GoHome}) if the alarm cannot be sounded within 2~seconds. \hfill $\Box$
\end{example}

\begin{lstlisting}[language = SLEEC, label={lst:trigger2}, caption={Extended version of \textsf{Rule2} with \kw{otherwise} construct},captionpos=b, float=tp, frame = single]
  Rule2_a when CameraStart and personNearby 
          then SoundAlarm within 2 seconds 
          otherwise GoHome
\end{lstlisting}

\noindent%
Thus, the \kw{otherwise} construct allows us to provide a different response, in the particular case of a timeout arising from the definition of a related \kw{within}. 

Another form of \textbf{constraint} requires an event \kw{not} to happen. In this case, a time budget must be defined via the \kw{within} construct, so as to not permanently disable the event.  

\begin{example}
  \textsf{Rule3} from Listing~\ref{lst:trigger} is triggered when \textsf{SoundAlarm} happens. In this case, for social reasons, the ``output'' event \textsf{GoHome} is blocked for 5~minutes. It may be the case, for example, that the teleoperators are in the home region, and the UAV should not come close to them while the alarm is sounding.  \hfill $\Box$
\end{example}

\noindent
The environment in which an autonomous agent is deployed is generally highly complex and the assumptions that underpin SLEEC rules may be invalid under certain conditions. To support resilience in such environments, we allow for the use of defeasible reasoning when there are scenarios leading to reasons that outweigh or disable a constraint~\cite{Brunero22DefeasingL}.  Defeasible reasoning is supported in our SLEEC language via \kw{unless} clauses, which allow normative rules to be be modified in light of additional information obtained from measures. 

\begin{example}
Listing~\ref{lst:trigger} presents a \textsf{Rule4} for constraining \textsf{CameraStart} with a view different from that in the previous rules. % in Listings~\ref{lst:trigger} and \ref{lst:trigger2}.
In \textsf{Rule4}, \textsf{CameraStart} is required to lead to \textsf{SoundAlarm}.  We have, however, an \kw{unless} clause with a condition depending on the value of the Boolean measure \textsf{personNearby}. If this measure is \textsf{true}, then \textsf{Rule4} requires the UAV to \textsf{GoHome}, so as to avoid the anti-social action of sounding an alarm near a person, likely a human firefighter.    
This is, however, once again defeated by a second \kw{unless} clause based on the \textsf{temperature} measure. If this measure is greater than 35$^{\circ}$C, no response is required. That is because such a high temperature is deemed an indication that there is a fire nearby, which trumps the legal/ethical concerns about filming a bystander, and the firefighter UAV is permitted to use its camera without restrictions. \hfill $\Box$
\end{example}

\noindent
Overall, multiple defeaters (grouped, if needed, within curly brackets $\{\ldots\}$ to indicate the constraint they apply to) alongside time constrains and timeouts can be defined in SLEEC rules. So, the semantics of the rules (formalised in the next section) can be rather subtle, and the interactions between multiple rules can be unexpected.

%% file: semantics.tex
%%%%%%%%%%%%%%%%%%%%%%%%%%%%%%%%%%%%%%%%%%%%%%%%%%%%%%%%%%%%
%%%%%%%%%%%%%%%%%%%%%%%%%%%%%%%%%%%%%%%%%%%%%%%%%%%%%%%%%%%%

\input{csp-table.tex}

%%%%%%%%%%%%%%%%%%%%%%%%%%%%%%%%%%%%%%%%%%%%%%%%%%%%%%%%%%%%

\input{semantic-rules.tex}

%%%%%%%%%%%%%%%%%%%%%%%%%%%%%%%%%%%%%%%%%%%%%%%%%%%%%%%%%%%%

This section defines the semantics of SLEEC using the process algebra $tock$-CSP, a timed variant of CSP~\cite{Ros98}. CSP is part of a large family of notations for specifying concurrent systems~\cite{Mil83,Mil99,BK85}, and is distinctive in its denotational semantics, giving rise to notions of refinement useful for stepwise development. A powerful model checker called FDR~\cite{FDR} supports the validation and verification of CSP (and $tock$-CSP) specifications. 

In Section~\ref{section:csp}, we give a brief introduction to $tock$-CSP. Section~\ref{section:semantics-overview} gives an overview of our semantics, with an example.  The detailed semantics of SLEEC triggers and responses are described in Sections~\ref{section:semantics-triggers} and \ref{section:semantics-responses}, respectively. 
%%%%%%%%%%%%%%%%%%%%%%%%%%%%%%%%%%%%%%%%%%%%%%%%%%%%%%%%%%%%
\subsection{Overview of $tock$-CSP}
\label{section:csp}
%%%%%%%%%%%%%%%%%%%%%%%%%%%%%%%%%%%%%%%%%%%%%%%%%%%%%%%%%%%%
CSP processes specify patterns of interaction via synchronisation on channels, taking into account (non)de\-ter\-minism, deadlock, and termination. Communications between parallel processes and with the environment are achieved via channels. These communications are instantaneous, atomic CSP events, that can carry values:~inputs and outputs. The dialect $tock$-CSP, in
addition, allows processes to specify time budgets and deadlines using a
special CSP event called $tock$. 

In Table~\ref{table:csp-operators} we summarise the $tock$-CSP operators that
we use in this paper. To illustrate the notation we present a $tock$-CSP process for a UAV firefighter's autopilot. 
\begin{example}\setlength{\zedindent}{0pt} 
  In this example, we define a process $AP$ to model a simple autopilot.  We use events $Navigate$ and $Track$ to represent capabilities of the drone to move to an area of interest~($Navigate$) and then search~($Track$) a fire. In $AP$, we specify that the autopilot first accepts a request to $Navigate$ and then~($\then$) starts $Track$ing. When a fire is found, $AP$ behaves as defined in the process $FIRE$. When $FIRE$ terminates, in sequence~($\cseq$) $AP$ recurses.
  \begin{displaymath}
    AP = Navigate \then Track \then FIRE \cseq AP
    \\
    FIRE = 0 \startBy (temperature?t \then 
    \\
    \t3
      \begin{array}[t]{l}
        \cspkw{if}~t > 35
        \\
        \cspkw{then}~1 \startBy (SoundFireAlarm \then \cspkw{Skip})
        \\
        \cspkw{else}~\cspkw{Skip})
      \end{array}  
  \end{displaymath}
  In $FIRE$ the autopilot reads a value $t$ using a channel $temperature$~($temperature?t$), and then behaves as defined by a conditional.  The process with the communication $temperature?t$ follow by the conditional is the argument of the operator~($\startBy$) that defines a deadline, here 0, for that process to exhibit visible behaviour.
  %:~either an interaction or termination. 
  The deadline defines the number of time units that can pass, that is, the number of $tock$ events that can occur, before the visible behaviour happens.  With the deadline 0, we specify that the input must happen immediately:~no $tock$ events are allowed before the communication on the channel $temperature$ occurs. In the conditional, if the temperature read ($t$) is greater than 35 Celsius, then an event $SoundFireAlarm$ is required to happen in at most 1~time unit. So, $SoundFireAlarm$ can happen before a $tock$ or after at most one $tock$. Afterwards, $FIRE$ terminates~($\cspl{Skip}$) immediately:~no more $tock$ events can happen.  If $t$ is less than or equal to 35, $FIRE$ just terminates. \hfill $\Box$
\end{example}
\noindent
We note that the CSP event $temperature$ corresponds to the SLEEC measure \textbf{temperature} in Listing~\ref{lst:trigger}. The other events are not mentioned there. In general, we can expect SLEEC rules to be concerned with some, but not all, capabilities of an agent.  Verification needs to take that into account, as we discuss in Section~\ref{subsec:verification}.

%%%%%%%%%%%%%%%%%%%%%%%%%%%%%%%%%%%%%%%%%%%%%%%%%%%%%%%%%%%%
\subsection{Overview of SLEEC semantics}
\label{section:semantics-overview}
%%%%%%%%%%%%%%%%%%%%%%%%%%%%%%%%%%%%%%%%%%%%%%%%%%%%%%%%%%%%

The semantics of a SLEEC \textbf{specification} from Figure~\ref{lst:BNF} is given by a function $\lsem\_\rsemS$ defined in Table~\ref{table:semantics}. This function maps the \textbf{specification} to a $tock$-CSP process, and is defined in terms of two other functions, $\lsem\_\rsemDS$ and $\lsem\_\rsemRS$, that capture the semantics of the definitions \textsf{dB} and rules \textsf{rB} of the \textbf{specification}.  The definitions in Table~\ref{table:semantics} are mechanised in our SLEEC tool~\cite{YBJCC23}, which automates the generatation of the $tock$-CSP semantics of a SLEEC \textbf{specification}~(see Section~\ref{section:tool}).

The semantics of \textbf{definitions} from Figure~\ref{lst:BNF} is given by corresponding declarations of channels and constants representing the SLEEC events, measures, and constants.  
The types \kw{boolean} and \kw{numeric} are given semantics as \cspkw{Bool} and \cspkw{Int}. For a \kw{scale} type, the semantics is a CSP \cspkw{datatype} that declares its literal parameters, and an associated Boolean function to record the order between those literals. A SLEEC constant becomes a CSP constant. 

\begin{example}
In Figure~\ref{fig:definitions-semantics-example}, we present the declarations for the definitions in Listing~\ref{lst:defBlock}. For each event and measure, we have a \cspkw{channel} declaration. For the type of the measure \textsf{windSpeed}, we define a \cspkw{datatype} $\cspl{STwindSpeed}$ and a Boolean function $\cspl{STlewindSpeed}$ with arguments $\cspl{v1windSpeed}$ and $\cspl{v2windSpeed}$ (of type $\cspl{STwindSpeed}$). If $\cspl{v1windSpeed}$ is the first literal \textsf{light}, then it is guaranteed to be less than or equal to $\cspl{v2windSpeed}$, no matter the value of $\cspl{v2windSpeed}$. If, however $\cspl{v1windSpeed}$ is \textsf{moderate}, then the inequality holds if $\cspl{v2windSpeed}$ is not \textsf{light}, since it is then at least \textsf{moderate} as well. Finally, if $\cspl{v1windSpeed}$ is \textsf{strong}, then the inequality holds if, and only if, $\cspl{v2windSpeed}$ is \textsf{strong} too. For model checking, we need to define a value for the constants. In this example, we use 3 as a time unit for the value of \textsf{ALARM\_DEADLINE}. \hfill $\Box$
\end{example}

\begin{figure}[t]
  \begin{displaymath}
    \cspkw{channel}~BatteryCritical
    \\
    \cspkw{channel}~CameraStart
    \\
    \cspkw{channel}~SoundAlarm
    \\
    \cspkw{channel}~GoHome
    \\
    \cspkw{channel}~personNearby: \cspkw{Bool}
    \\
    \cspkw{channel}~temperature: \cspkw{Int}
    \\
    \cspkw{channel}~windSpeed: STwindSpeed
    \\
    \cspkw{datatype}~STwindSpeed = light | moderate | strong
    \\
    STlewindSpeed(v1windSpeed,v2windSpeed) = 
    \\
    \quad
      \begin{array}[t]{l}
      \cspkw{if}~v1windSpeed == light 
      \\
      \cspkw{then}~\cspkw{true}
      \\
      \cspkw{else}~(\begin{array}[t]{l}
                      \cspkw{if}~v1windSpeed == moderate
                      \\
                      \cspkw{then}~(v2windSpeed \notin \{light\})
                      \\
                      \cspkw{else}~v2windSpeed == strong~~~)
                    \end{array}  
      \\
    \end{array}
    \\
    ALARM\_DEADLINE = 3 
  \end{displaymath}    
  \caption{\label{fig:definitions-semantics-example}CSP declarations for the definitions in Listing~\ref{lst:defBlock}}  
\end{figure}

\noindent
The recursive definition of $\lsem\_\rsemDS$ is given by two equations.  For a single definition \textsf{def}, the semantics is given by another function $\lsem \_\rsemD$.  For a list of definitions \textsf{def defS}, containing a single definition \textsf{def} followed by a list \textsf{defS}, the semantics is the sequence of CSP declarations determined by $\lsem\textsf{def}\rsemD$ to capture the semantics of \textsf{def}, followed by the CSP declarations defined by a recursive application of $\lsem\_\rsemDS$ to \textsf{defS}. We do not consider the empty list of definitions, since a SLEEC specification defines restrictions on the use of the capabilities of the agent, and without a declaration of capabilities, there is no sensible specification. 

The equations defining $\lsem\_\rsemD$ consider each form of \textbf{definition} in turn. We assume that identifiers in SLEEC satisfy the usual lexical restrictions adopted in CSP, so that, for example, events and measures are represented by channels of the same name. For a \kw{constant}, we assume that a value is given. The type used in the declaration of a measure is given by the semantic function $\lsem\_ \rsemT$ whose arguments are the type \textsf{T} and the identifier \textsf{mID} of the measure.  The equations defining $\lsem\_ \rsemT$ for \kw{boolean} and \kw{numeric} are straightforward. For a \kw{scale} type, we use the name of the measure in defining the corresponding CSP declarations. 

 For simplicity, we use an informal notation to represent a \kw{scale} type with $n$ parameters $sp_1$ to $sp_n$, namely, $\kw{scale}(\meta{sp_1, \ldots, sp_n})$.  The definition of a formal generative function for the semantics of a measure with such a type is, however, straightforward. The name of the \cspkw{datatype} defined is that of the measure, that is, the argument \textsf{mID}, prefixed with $ST$.  The name of the Boolean function that defines its order is prefixed with $STle$ instead.  

The recursive definition of the semantic function $\lsem \_\rsemRS$ for a list of rules is similar to that of $\lsem\_\rsemDS$, but is based on the semantic function $\lsem\_\rsemR$ for a \textbf{rule}. The semantics of each rule is given by a process, named after that rule, and defined using two processes that capture the meaning of its \textbf{trigger} and of its \textbf{response}. 
The process for every rule is defined by composing in sequence a $Trigger$ and a $Monitoring$ process. This reflects the fact that a rule imposes no constraints until its trigger is observed. At that point, it monitors~(that is, determines) the allowed behaviour to enforce the response.  

\begin{example}
    For \textsf{Rule2} in Listing~\ref{lst:trigger}, the CSP process that defines its semantics is shown in Figure~\ref{fig:rule2}. The behaviour of the process $Rule2$ is initially defined by that of $TriggerRule2$.  When the trigger of \textsf{Rule2} is observed, $TriggerRule2$ terminates (via the $\cspkw{Skip}$ from the conditional statement), and the process $MonitoringRule2$ takes over.  When the response happens, $MonitoringRule2$ terminates and $Rule2$ recurses. \hfill $\Box$
\end{example}

\begin{figure}\setlength{\zedindent}{0pt}
    \begin{displaymath}
        Rule2 = TriggerRule2 \cseq MonitoringRule2 \cseq Rule2
        \\
        TriggerRule2 = 
        \\
        \quad
          \begin{array}{l}
            \cspkw{let}~MTrigger = 0 \startBy (personNearby?vpersonNearby \then
            \\
            \t1
              \begin{array}{l}
                \cspkw{if}~vpersonNearby == true
                \\
                \cspkw{then}~\cspkw{Skip}
                \\
                \cspkw{else}~TriggerRule2)
              \end{array}  
            \\
            \cspkw{within}~\begin{array}[t]{l}
                             CameraStart \then MTrigger
                             \\
                             \extchoice
                             \\
                             SoundAlarm \then TriggerRule2
                           \end{array}  
        \end{array}
        \\
        MonitoringRule2 = 2 \startBy (SoundAlarm \then \cspkw{Skip})
    \end{displaymath}
    \caption{Semantics of \textsf{Rule2} in Listing~\ref{lst:trigger}}
    \label{fig:rule2}
\end{figure}

\noindent%
In Table~\ref{table:semantics}, the definition of $\lsem\_\rsemR$ uses the identifier $\textsf{rID}$ of the rule to assemble the identifiers of the $Trigger$ and $Monitoring$ processes.  The definition of the $Trigger$ process is given by the semantic function $\lsem\_\rsemTG$ whose arguments are the \textsf{trig}ger of the rule, the alphabet of events, that is, the set of all events used in the response of the rule, and two continuation processes. The alphabet of events is given by the function $\meta{\alpha_E(resp)}$. The first continuation process determines the behaviour when the trigger happens. In the definition of $\lsem\_\rsemR$, this is \cspkw{Skip}, since the $Trigger$ process must terminate in this case. The second continuation process determines the behaviour if the event of the trigger takes place, but its condition does not hold.  In the definition of $\lsem\_\rsemR$, this is the $Trigger$ process itself, since in this case the $Trigger$ process must recurse. 

To define the $Monitoring$ process, we use the semantic function $\lsem\_\rsemRDS$. The first argument is the \textbf{response} that is to be monitored.  The subsequent arguments are needed because a \textbf{defeater} may void the monitoring, which then needs to check the trigger again.  The extra arguments are the \textsf{trig}ger and the alphabet of events used in the response.
The final argument of $\lsem\_\rsemRDS$ is a continuation process, which in the definition of $\lsem\_\rsemR$ is $Monitoring$. 

We define $\lsem\_\rsemTG$ next, and $\lsem\_\rsemRDS$ in Section~\ref{section:semantics-responses}. Those familiar with the use of CSP to specify properties might observe that we do not adopt the usual approach that considers the overall alphabet of events, and defines a rule that imposes no restrictions outside its own alphabet.  That approach is convenient for verification by refinement, but does not easily support checks for conflicts and redundancy.  With our semantics, we support validation, and, for verification, we adopt a more elaborate notion of correctness, using refinement and priorities (cf.~Section~\ref{sec:validation}). 

%%%%%%%%%%%%%%%%%%%%%%%%%%%%%%%%%%%%%%%%%%%%%%%%%%%%%%%%%%%%
\subsection{Triggers}
\label{section:semantics-triggers}
%%%%%%%%%%%%%%%%%%%%%%%%%%%%%%%%%%%%%%%%%%%%%%%%%%%%%%%%%%%%

The definition of $\lsem\_\rsemTG$ is given in Table~\ref{table:semantics-triggers}. For a \textbf{trigger} that has just an event \textsf{eID}, the process is a synchronisation on that event followed by the argument process \textsf{sp} that defines the continuation when the trigger happens. A choice allows the response events to happen freely, but their occurrence leads to a recursion so that the rule is not enforced if the trigger has not happened. 

\input{semantics-triggers}

If the \textbf{trigger} has a Boolean expression on measures, the process is defined using \cspkw{let} and \cspkw{within} clauses. The actual process is defined in the \cspkw{within} clause, but in its definition we can use processes named in the \cspkw{let} clause.  In the process $\lsem \textsf{eID}~\kw{and}~\textsf{mBE},\textsf{AR},\textsf{sp},\textsf{fp}\rsemTG$, we have the synchronisation on \textsf{eID} is followed by a process $MTrigger$, defined in the \cspkw{let} clause using a semantic function $\lsem \_\rsemME$.

\begin{example}
    As shown in Figure~\ref{fig:rule2}, if the trigger has a Boolean expression on measures, $MTrigger$ first reads the values of the measures urgently. For \textsf{Rule2} in Listing~\ref{lst:trigger}, the condition is just on the measure \textsf{personNearby}, so $MTrigger$ inputs a value $vpersonNearby$ using the channel $personNearby$. Afterwards, a conditional checks the measure expression. If it holds, the trigger has occurred, and $MTrigger$  terminates, leading to $TriggerRule2$ terminating as well. Otherwise, $MTrigger$ recurses back to the $Trigger$ process to wait for the trigger event again. \hfill $\Box$
\end{example}

\noindent%
The function $\lsem\_\rsemME$ takes the list of measures used in the Boolean expression as arguments, that measure condition itself, and the continuation processes. In the definition of $\lsem \_ \rsemTG$, the first argument $\meta{\alpha_{ME}(mBE)}$ of $\lsem\_\rsemME$ is defined by a function $\meta{\alpha_{ME}}$, similar to $\meta{\alpha}$, but providing just measure identifiers used in the Boolean expression. 

The inductive definition of $\meta{\lsem\_\rsemME}$ considers separately an empty list $\lseq\rseq$ of measures and a list with at least one measure $\meta{mID}$. In the process defined by this function, the value $v\meta{mID}$ of each measure $\meta{mID}$ is read urgently in sequence~($\then$). That value is then substituted for $\meta{mID}$ in the expression $\meta{mBE}$.  Once all of the measures are input, a conditional checks the value of the resulting expression. 

In detail, for a list of identifiers $\meta{\lseq mID\rseq \cat mIDs}$, the definition of $\lsem\_\rsemME$ defines a process that reads the value of $\meta{mID}$ and records it into a local variable named $v\meta{mID}$. To make that urgent, it uses the operator $\startBy$ with deadline 0 over a process that starts with the communication $\meta{mID}?v\meta{mID}$. The behaviour that follows is defined by the process characterised by a recursive application of $\lsem\_\rsemME$. 

In that application of $\lsem\_\rsemME$, the remaining measures in $\meta{mIDs}$ are considered. Moreover, the Boolean expression is changed to refer to the variable $v\meta{mID}$, where the measure $\meta{mID}$ is used. We use $\meta{mBE[}v\meta{\meta{mID}/\meta{mID}]}$ to denote the Boolean expression obtained by replacing the occurrences of $\meta{mID}$ with $v\meta{mID}$.  In our example semantics for \textsf{Rule2} in Listing~\ref{lst:trigger}, \textsf{personNearby} becomes $vpersonNearby$.

If the first argument of $\lsem\_\rsemME$ is the empty list of measures, then all the relevant measure values have been read, and the Boolean expression is defined in terms of those values~(recorded in local variables).  So, $\lsem\_\rsemME$ defines a conditional process that specifies the appropriate continuation behaviour depending on the measure condition.  

The actual condition evaluated is specified using a normalisation function. In  $\meta{norm(mBE)}$ the SLEEC relational operators applied to literals of \kw{scale} types in $\meta{mBE}$ are encoded using the comparator functions of those \kw{scale} types.  (Strictly speaking, $\meta{norm(\_)}$ requires an extra argument defining the type of the measures and the names of the comparator functions for the \kw{scale} types.) Additionally, the use of a measure $mID$ as a Boolean is transformed to an equality $mID == \cspkw{true}$ as required by the CSP notation.

%%%%%%%%%%%%%%%%%%%%%%%%%%%%%%%%%%%%%%%%%%%%%%%%%%%%%%%%%%%%
\subsection{Responses}
\label{section:semantics-responses}
%%%%%%%%%%%%%%%%%%%%%%%%%%%%%%%%%%%%%%%%%%%%%%%%%%%%%%%%%%%%

\input{semantics-responses}

The semantic function $\lsem\_\rsemRDS$ for \textbf{response} definitions is specified in Table~\ref{table:semantics-responses}.  The semantics of a \textbf{response} enclosed in curly brackets is just the semantics of its \textbf{constraint} and \textbf{defeaters} itself. We omit that simple definition from Table~\ref{table:semantics-responses}.  The semantics of a response that has just a constraint is given by the function $\lsem\_\rsemRP$. 

\begin{example}
    \textsf{Rule2} in Listing~\ref{lst:trigger} provides an example of a response that has just a constraint, that is, the rule contains no defeaters. The $Monitoring$ process in its semantics, shown in Figure~\ref{fig:rule2}, captures the time constraint in the response. It requires that $SoundAlarm$ takes place within 2 time units. In this case, we have the assumption that each $tock$ represents the passage of 1~second. \hfill $\Box$
\end{example}

\noindent%
The SLEEC rules can refer to a variety of time units. To give semantics, we can either assume that $tock$ represents the passage of a minimal period of time that can be considered~(1 millisecond, for example), or calculate the greatest common divisor of all periods of time referenced, and adopt that to define the meaning of $tock$.  Whatever the solution, when using a time period definition we need to normalise the value to describe it in terms of a number of $tock$ events. For instance, if $tock$ is deemed to represent a second, then ``1 minute'' should be normalised to 60.  

The definition of $\lsem\_\rsemRP$ has one equation for each possible form of constraint.  If it is just an event, the constraint process defined by $\lsem\_\rsemRP$ requires that event to be accepted and then terminates.  We recall that termination indicates that the constraint has been satisfied, and the rule process can recurse and wait for the next trigger.

If there is a time budget $\meta{\kw{within}~v~tU}$ defining a number $\meta{v}$ of time units given by $\meta{tU}$, then the process for the event is included in an $\startBy$. The deadline $\meta{norm(v,tU)}$ is determined using a normalisation function to calculate the number of $tock$ events allowed, as explained above. 

If the possibility of a timeout is considered, via an \kw{otherwise} clause, instead of a $\startBy$, we use a timed interrupt~($\interrupt_d$) to specify that, if the budget $\meta{norm(v,tU)}$ is used up, the process that captures the semantics of the response associated with the \kw{otherwise} takes over. 

\begin{example}%\setlength{\zedindent}{0pt}
    The $MonitoringRule2\_a$ process for the SLEEC \textsf{Rule2\_a} in Listing~\ref{lst:trigger2} is shown below.
    \begin{displaymath}
      (SoundAlarm \then \cspkw{Skip}) \interrupt_2 (GoHome \then \cspkw{Skip})
    \end{displaymath}
    If after 2 seconds, for whatever reason, the alarm cannot be sounded, then $GoHome$ is required.
    \hfill $\Box$
\end{example}

\noindent%
Finally, for a constraint that forbids the occurrence of an event, the semantics is the process $\meta{\cspkw{Wait}(norm(v,tU))}$ that pauses:~only allows time to pass, that is, $tock$ events to happen, for $\meta{v~tU}$ time units, and then terminates.

If a response has one or more defeaters, $\lsem\_\rsemRDS$ defines a process using \cspkw{let} and \cspkw{within} clauses.  The \cspkw{let} clause defines a number of local $Monitoring$ processes used in the \cspkw{within} clause to define the overall $Monitoring$ process.

\begin{example}\label{example:rule4a-semantics}
  Consider the simpler version of \textsf{Rule4} in Listing~\ref{lst:trigger3}. Its $Monitoring$ process is as follows. 
  \begin{displaymath}
      \cspkw{let} 
      \\
      \quad 
        \begin{array}{l}
          Monitoring1 = SoundAlarm \then \cspkw{Skip}
          \\
          Monitoring2 = GoHome \then \cspkw{Skip}
        \end{array}
      \\
      \cspkw{within}
      \\
      \quad
      \begin{array}{l}
        0 \startBy (personNearby?vpersonNearby \then 
        \\
        \quad
          \begin{array}{l}
            \cspkw{if}~vpersonNearby == \cspkw{true}
            \\
            \cspkw{then}~Monitoring2~\cspkw{else}~Monitoring1)
          \end{array}
      \end{array} 
  \end{displaymath}
  The constraint in the \kw{then} clause and the response in the \kw{unless} clause are captured by local processes $Monitoring1$ and $Monitoring2$. In the \cspkw{within} clause, after reading the relevant measures, the process chooses a local $Monitoring$ process based on the \kw{unless} condition. \hfill $\Box$
\end{example}

\begin{lstlisting}[language = SLEEC, label={lst:trigger3}, caption={Simpler version of \textsf{Rule4}},captionpos=b, float=tp, frame = single]
  Rule4_a when CameraStart then SoundAlarm 
          unless personNearby then GoHome 
\end{lstlisting}

\noindent%
The local $Monitoring$ processes are defined by $\lsem\_\rsemLRDS$, which takes as argument a list containing the \textsf{const}raint of the response, and the responses in the defeaters \textsf{dfts}. We use $\meta{dfts\filter_{RP}}$ to represent the list of those responses. For an \kw{unless} defeater without a response, we get $\cspkw{NoRep}$. This is a special response defined in the semantics just for the purposes of simplifying the semantic rules.  The additional arguments of $\lsem\_\rsemLRDS$ are the extra arguments of $\lsem\_\rsemRDS$, and a counter for the $Monitoring$ processes used to define their names.  In the definition of $\lsem\_\rsemRDS$, we define the (initial) value of the counter as 1. 

The definition of $\lsem\_\rsemLRDS$ has two equations for a singleton list of responses, and a third equation for a list $\meta{\lseq resp\rseq \cat resps}$ starting with a response \textsf{resp} followed by a list \textsf{resps}.  For a singleton list with a proper response \textsf{resp}, the $Monitoring$ process is defined by $\lsem\_\rsemRDS$.  For the special response \cspkw{NoRep}, we need to consider the trigger.

\begin{example}\setlength{\zedindent}{0pt}\label{example:rule4-semantics}
  The $Monitoring$ process for \textsf{Rule4} in Listing~\ref{lst:trigger3} is shown below. It is similar to that in Example~\ref{example:rule4a-semantics}, but has an extra local $Monitoring3$ process since we have an extra \kw{unless} clause. In the \cspkw{within} clause, both relevant measures are read urgently, and then conditionals identify the right local process to monitor the behaviour. 
  \begin{displaymath}
      \cspkw{let} 
      \\
      \quad 
        \begin{array}{l}
          Monitoring1 = SoundAlarm \then \cspkw{Skip}
          \\
          Monitoring2 = GoHome \then \cspkw{Skip}
          \\
          Monitoring3 = \begin{array}[t]{l}
                          CameraStart \then MonitoringRule4
                          \\
                          \extchoice SoundAlarm \then Monitoring3
                          \\
                          \extchoice GoHome \then Monitoring3
                          % \\
                          % \extchoice personNearby?x \then Monitoring3
                          % \\
                          % \extchoice temperature?x \then Monitoring3
                        \end{array}  
        \end{array}
      \\
      \cspkw{within}
      \\
      \quad
      \begin{array}{l}
        0 \startBy (personNearby?vpersonNearby \then 
        \\
        \quad
          \begin{array}{l}
            0 \startBy (temperature?vpersonNearby \then
            \\
            \quad
            \begin{array}{l}
              \cspkw{if}~vtemperature > 35
              ~%\\
              \cspkw{then}~Monitoring3
              \\
              \cspkw{else}~(\cspkw{if}~vpersonNearby == \cspkw{true}
              \\
              \cspkw{then}~Monitoring2~\cspkw{else}~Monitoring1)))
            \end{array}  
          \end{array}
      \end{array} 
  \end{displaymath}
  $Monitoring3$ corresponds to the \kw{unless} clause for the condition \textsf{temperature $>$ 35}, which does not have an associated response. The meaning in this case is that the rule imposes no restrictions. So, all CSP events used in the rule semantics need to be allowed.  In the example, the $Monitoring3$ process needs to offer the choice~($\extchoice$) of all CSP events corresponding to the events used in \textsf{Rule4}.  For the event in the trigger, its occurrence leads to the outer $MonitoringRule4$ process taking over.  For all other events, $Monitoring3$ simply recurses:~$Monitoring3$ does not block these events, but ignores them by just proceeding. 
\end{example}
\noindent%
As illustrated, the definition of a local $Monitoring$ process for an \kw{unless} clause without a response, that is, with a response \cspkw{NoRep}, requires information about the trigger of the overall rule, its alphabet, and the overall $Monitoring$ process.  These are the extra arguments of $\lsem\_\rsemRP$ and $\lsem\_\rsemLRDS$. 

For a response \cspkw{NoRep}, the local process $Monitoring\meta{n}$ defined by $\lsem\_\rsemLRDS$, when applied to a counter value $\meta{n}$, is specified as a choice over a trigger process characterised by $\lsem\_\rsemTG$ and a choice of events $e$ from the alphabet $\meta{AR}$ of the rule given as argument.  In every choice, such an event $e$ is followed by a recursion.  In our example above, $\meta{AR}$ contains $SoundAlarm$ and $GoHome$. The arguments of $\lsem\_\rsemTG$ are the $\meta{trig}$ger, the process $\meta{mp}$, providing the continuation in case the trigger occurs, and the process $Monitoring\meta{n}$, the continuation if the trigger does not occurs.  We recall that $\lsem\_\rsemR$ defines $\meta{mp}$ as the $Monitoring$ process for the rule. 

The process in the \cspkw{within} clause of a response process~(as defined by $\lsem\_\rsemRDS$) is specified by a $\lsem\_\rsemCDS$ function. Its arguments are the alphabet $\alpha_{ME}\meta{(dfts)}$ of measures of the defeaters $\meta{dfts}$, the defeaters $\meta{dfts}$ themselves, and the number of responses to be handled, namely, the number $\# dfts$ of defeaters, plus 1, to consider the $\meta{const}$raint in the rule overall response.  The definition of $\alpha_{ME}\meta{(dfts)}$ is similar to that of $\alpha_{ME}\meta{(mBE)}$, but applies to a list of defeaters, considering the Boolean expressions that they use.  

The inductive definition of $\lsem\_\rsemCDS$ is simple.  For a list of measures $\meta{\lseq mID\rseq \cat mIDs}$, the process inputs the values $v\meta{mID}$ or the measure urgently and then behaves as the process defined by $\lsem\_\rsemCDS$ for $\meta{mIDs}$. In the defeaters used as argument for the recursive application of $\lsem\_\rsemCDS$, the references to $\meta{mID}$ are replaced with $v\meta{mID}$. In Example~\ref{example:rule4-semantics}, this defines the two urgent communications to input values of the measures \textsf{personNearby} and \textsf{temperature}.

For an empty list of measures, the process is defined by $\lsem\_\rsemEDS$.  This is again an inductive definition, whose arguments are the list of defeaters, the local $Monitoring1$ process that applies when no defeater in the list does, and the number of defeaters in that list.  We recall that $Monitoring1$ is the process that captures the behaviour of the overall constraint of the rule~(see definitions of $\lsem\_\rsemRDS$ and $\lsem\_\rsemLRDS$). If the list of defeaters $\meta{dfts~dft}$ has more than than one defeater, the result is the application of $\lsem\_\rsemEDS$ to the last defeater, whose monitoring process is given by a recursive application of $\lsem\_\rsemEDS$ to define a process for the other defeaters, which applies when $\meta{dft}$ does not.   

If we have just one defeater, then the process is a conditional that checks whether its condition applies. It it does, the local $Monitoring$ process identified by the counter $\meta{n}$ is used. Otherwise, the continuation process $\meta{fp}$ is used.  

\begin{example}\setlength{\zedindent}{0pt}
    We show below the use of $\lsem\_\rsemEDS$ to define the conditional in Example~\ref{example:rule4-semantics}.
    \begin{argue}
        \lsem \begin{array}[t]{l}
                \kw{unless}~vpersonNearby~\kw{then}~\meta{GoHome}
                \\
                \kw{unless}~vtemperature > 35, 
                \\
                Monitoring1, 3~~\rsemEDS
              \end{array}  
        \\
        = 
        \\
          \lsem \begin{array}[t]{l}
                  \kw{unless}~vtemperature > 35,
                  \\
                  \lsem \kw{unless}~vpersonNearby~\kw{then}~\meta{GoHome}, Monitoring1, 2 \rsemEDS, 
                  \\ 
                  3~~\rsemEDS
                \end{array} 
        \\
        = 
        \\
          \lsem \begin{array}[t]{l}
                  \kw{unless}~vtemperature > 35,
                  \\
                  (\begin{array}[t]{l}
                     \cspkw{if}~vpersonNearby == true
                     \\
                     \cspkw{then}~Monitoring2~\cspkw{else}~Monitoring1~~),
                   \end{array}  
                  \\ 
                  3~~\rsemEDS
                \end{array} 
        \\
        = 
        \\
        \begin{array}[t]{l}
          \cspkw{if}~vtemperature > 35
          \\
          \cspkw{then}~Monitoring3
          \\
          \cspkw{else}~
          (\begin{array}[t]{l}
             \cspkw{if}~vpersonNearby == true
             \\
             \cspkw{then}~Monitoring2~\cspkw{else}~Monitoring1~~)
           \end{array}  
        \end{array} 
    \end{argue}
\end{example}
\noindent%
Additional examples are provided in the next sections.  \hfill $\Box$

%% file: csp-table.tex
\def\arraystretch{1.2}
\begin{table*}[t]\centering
    \caption{\label{table:csp-operators} List of $tock$-CSP operators, with basic processes at the top, followed
    by composite processes:~{$\cspl{P}$} and {$\cspl{Q}$} are metavariables
    that stand for processes, $\cspl{d}$ for a numeric
    expression, {$\cspl{e}$} for an event, {$\cspl{a}$} and {$\cspl{c}$} for
    channels, {$\cspl{x}$} for a variable, {$\cspl{I}$} for a set,
    {$\cspl{v}$} for an expression, {$\cspl{g}$} for a condition, and
    {$\cspl{X}$} for a set of events. For a channel {$\cspl{c}$},
    {$\cspl{\lchan c \rchan}$} is a set of events; if $\cspl{c}$ is a typed
    channel then events are constructed using the dot notation, so that
    $\cspl{\lchan c \rchan = \lchan c.v_0, ..., c.v_{n}\rchan}$, where
    $\cspl{v_{i}}$ ranges over the type of $\cspl{c}$.
    } 
    \begin{tabular}{lp{14cm}}
        \toprule
        \textbf{Process} & \textbf{Description}\\
        \midrule
        {$\cspkw{Skip}$} & \textbf{Termination}: terminates immediately\\
        %\hline
        {$\cspkw{Wait}\cspl{(d)}$} & \textbf{Delay}: terminates exactly after {$\cspl{d}$} units of time have elapsed\\
        % \hline
        % {$\cspl{Stop}$} & \textbf{Timed deadlock}: no events are offered, but time can pass\\
        % \hline
        % {$\cspl{Stop_{U}}$} & \textbf{Timelock}: no events are offered and time cannot pass\\
        % \hline
        % {$\cspl{\mu X @ P}$} & \textbf{Recursion}: defines a process $\cspl{X}$, that behaves like $\cspl{P}$, and where $\cspl{X}$ may occur in $\cspl{P}$\\
        % %Behaves like $\cspl{Stop}$, but in addition offers to synchronise on $\cspl{share}$ events indefinitely\\
        %\hline
        {$\cspl{e \then P}$} & \textbf{Prefix operator}: initially offers to engage in the event {$\cspl{e}$} while permitting any amount of time to pass, and then behaves as {$\cspl{P}$}\\
        %\hline
        {$\cspl{a?x \then P}$} & \textbf{Input prefix}: same as above, but offers to engage on channel {$\cspl{a}$} with any value, and stores the chosen value in {$\cspl{x}$}\\
        %\hline
        {$\cspl{a?x:I \then P}$} & \textbf{Restricted input prefix}: same as above, but restricts the value of {$\cspl{x}$} to those in the set {$\cspl{I}$}\\
        %\hline
        {$\cspl{a!v \then P}$} & \textbf{Output prefix}: same as above, but initially offers to engage on channel {$\cspl{a}$} with a value {$\cspl{v}$}\\
        %\hline
        {\cspkw{if}$\cspl{\; g\; }$\cspkw{then}$\cspl{\; P\; }$\cspkw{else}$\cspl{\; Q}$} & \textbf{Conditional}: behaves as {$\cspl{P}$} if the predicate {$\cspl{g}$} is true, and otherwise as {$\cspl{Q}$}\\
        %\hline
        {$\cspl{P \extchoice Q}$} & \textbf{External choice} of {$\cspl{P}$} or {$\cspl{Q}$} made by the environment\\
        % \hline
        % {$\cspl{P \intchoice Q}$} & \textbf{Internal choice} of {$\cspl{P}$} or {$\cspl{Q}$} made non-deterministically\\
        %\hline
        {$\cspl{P \cseq Q}$} & \textbf{Sequence}: behaves as {$\cspl{P}$} until it terminates successfully, and, then it behaves as {$\cspl{Q}$}\\
        %\hline
        {$\cspl{P \hide X}$} & \textbf{Hiding}: behaves like {$\cspl{P}$} but with all communications in the set {$\cspl{X}$} hidden\\
        %  \hline
        %  {\lstinline!P |\ X!} & Project: behaves like {\lstinline!P!} but with all communications not in the set {\lstinline!X!} hidden\\
        %\hline
        {$\cspl{P \interleave Q}$} & \textbf{Interleaving}: {$\cspl{P}$} and {$\cspl{Q}$} run in parallel and do not interact with each other\\
        %\hline
        {$\cspl{P \lpar X \rpar Q}$} & \textbf{Generalised parallel}: {$\cspl{P}$} and {$\cspl{Q}$} must synchronise on events that belong to the set {$\cspl{X}$},
            with termination occurring only when both {$\cspl{P}$} and {$\cspl{Q}$} agree to terminate\\
        % \hline
        % {$\cspl{P\lrename a \becomes c_{1}, ..., a \becomes c_{i}\rrename}$} & \textbf{Renaming}: replaces uses of channel {$\cspl{a}$} with channels {$\cspl{c_1}$} to {$\cspl{c_i}$} in {$\cspl{P}$}\\
        %\hline
        {$\cspl{P \interrupt Q}$} & \textbf{Interrupt}: behaves as {$\cspl{P}$} until an event offered by {$\cspl{Q}$} occurs, and then behaves as {$\cspl{Q}$}\\
        %\hline
        {$\cspl{P \interrupt_{d} Q}$} & \textbf{Strict timed interrupt}: behaves as {$\cspl{P}$}, and, after exactly {$\cspl{d}$} time units behaves as {$\cspl{Q}$}\\
        %\hline
        {$d \blacktriangleleft P$} & \textbf{Deadline for visible interaction}: engages in an event of $P$ in at most {$\cspl{d}$} time units
        \\
        % \hline
        % {$\cspl{P \exception{X} Q}$} & \textbf{Exception}: behaves as {$\cspl{P}$} until {$\cspl{P}$} performs an event in {$\cspl{X}$}, and, then behaves as {$\cspl{Q}$}\\
        % \hline
        % {$\cspl{\lpar ~X\rpar i : I @ P(i)}$} & \textbf{Replicated generalised parallel}: behaves as {$\cspl{P(i)}$} in parallel for all {$\cspl{i}$} in {$\cspl{I}$} synchronising in $X$\\
        % \hline
        % {$\cspl{\Interleave i : I @ P(i)}$} & \textbf{Replicated interleaving}: behaves as {$\cspl{P(i)}$} interleaved for all {$\cspl{i}$} in {$\cspl{I}$}\\
        %\hline
        {$\cspl{\Extchoice i : I @ P(i)}$} & \textbf{Replicated external choice}: offers an external choice over processes {$\cspl{P(i)}$} for all {$\cspl{i}$} in {$\cspl{I}$}\\
        % \hline
        % {$\cspl{\Intchoice i : I @ P(i)}$} & \textbf{Replicated internal choice}: offers an internal choice over processes {$\cspl{P(i)}$} for all {$\cspl{i}$} in {$\cspl{I}$}\\
        \bottomrule
        % {\lstinline!TRUN(X)!} & Continuously offers the events in the set {\lstinline!X!} to the environment, while time can pass\\
        % \hline
    %     {\lstinline!timed_priority(P)!} & Maximal progress: behaves as {\lstinline!P!} with internal behaviour given priority over $tock$, so that internal behaviour takes no time\\
    %   \hline
    \end{tabular}
\end{table*}

%% file: semantic-rules.tex
\begin{table*}\small
\caption{\label{table:semantics}Rules that define a tock-CSP semantics for SLEEC. We use the following {\em metavariables} in the definitions of the rules:~$\meta{def}$ as a metavariable to stand for an element of the syntactic category $\meta{definitions}$, $\meta{defS}$ to stand for an element of $\meta{definitions}$, $\meta{eID}$ for an $\meta{eventID}$, $\meta{mID}$ for a $\meta{measureID}$, $\meta{T}$ for a $\meta{type}$, $\meta{cID}$ for a $\meta{constID}$, $\meta{v}$ for a $\meta{value}$, $\meta{sp}$ and subscripted counterparts for a $\meta{scaleParams}$,$\meta{r}$ for a $\meta{rule}$, $\meta{rrS}$ for an element of $\meta{rules}$, $\meta{rID}$ for a $\meta{ruleID}$, $\meta{trig}$ for a $\meta{trigger}$, and finally $\meta{resp}$ for a $\meta{response}$. These metavariables are also used in rules in Tables~\ref{table:semantics-triggers} and \ref{table:semantics-responses}.}
\centering
\begin{tabular}{l@{\ } l@{\ } p{.42\textwidth}}
\toprule
%%%%%%%%%%%%%%%%%%%%%%%%%%%%%%%%%%%%%%%%%%%%%%%%%%%%%%%%%%%%%%%%%%%%%%%%%%%
% specification
%%%%%%%%%%%%%%%%%%%%%%%%%%%%%%%%%%%%%%%%%%%%%%%%%%%%%%%%%%%%%%%%%%%%%%%%%%%
$\lsem 
    % \begin{array}[t]{l}
    \kw{def\_start\ } 
      \meta{dB} 
    \kw{\ def\_end} \;\;%\\ 
    \kw{rule\_start\ }       
      \meta{rB} 
    \kw{\ rule\_end}~
  \rsemS
  % \end{array}
$
& = &
$\meta{~\lsem dB\rsemDS \quad \lsem rB\rsemRS}$
\\
\midrule
%
%%%%%%%%%%%%%%%%%%%%%%%%%%%%%%%%%%%%%%%%%%%%%%%%%%%%%%%%%%%%%%%%%%%%%%%%%%%
% definitions
%%%%%%%%%%%%%%%%%%%%%%%%%%%%%%%%%%%%%%%%%%%%%%%%%%%%%%%%%%%%%%%%%%%%%%%%%%%

%
%%%%%%%%%%%%%%%%%%%%%%
% definitions: single
%%%%%%%%%%%%%%%%%%%%%%
$\lsem
       \meta{def} 
   \rsemDS
$
& = &
$\lsem\meta{ def }\rsemD$
\\ %
% definitions: several
$\lsem
       \meta{def~defS} 
  \rsemDS
$
& = &
$\lsem \meta{def} \rsemD \quad \lsem \meta{defS} \rsemDS$
\\
%%%%%%%%%%%%%%%%%%%%
% definition: event
%%%%%%%%%%%%%%%%%%%%
$\lsem
       \kw{event~} \meta{eID} 
 \rsemD
$
& = &
$\cspkw{channel}~\meta{eID}$
\\ %
%%%%%%%%%%%%%%%%%%%%%%
% definition: measure
%%%%%%%%%%%%%%%%%%%%%%
$\lsem
       \kw{measure~} \meta{mID: T} 
 \rsemD
$
& = &
    $\cspkw{channel}~\meta{mID}: \lsem\meta{~T,mID\rsemT}$
\\ %
% definition: constant
$\lsem
       \kw{constant~} \meta{cID} = \meta{v} 
 \rsemD
$
& = &
$\meta{cID} = \meta{v}$
\\ %
% type: boolean
$\lsem
       \kw{boolean},\meta{mID}  
 \rsemT
$
& = &
$\cspkw{Bool}$
\\ %
%%%%%%%%%%%%%%%%%
% type: numeric
%%%%%%%%%%%%%%%%%
$\lsem
       \kw{numeric},\meta{mID}
 \rsemT
$
& = &
$\cspkw{Int}$
\\ %
%%%%%%%%%%%%%%%%%
% type: scale
%%%%%%%%%%%%%%%%%
$\lsem
       \kw{scale}(\meta{sp_1}, \ldots, \meta{sp_n}),\meta{mID} 
 \rsemT
$
& = &
$ST\meta{mID}$
\\
& &
$\cspkw{datatype}~ST\meta{mID} = \meta{sp_1} \mid \ldots \mid \meta{sp_n}$
\\
& & 
$\begin{array}{l}
   STle\meta{mID}(v1\meta{mID},v2\meta{mID}) = 
   \\
   \quad
     \begin{array}[t]{l}
       \cspkw{if~}v1\meta{mID} == \meta{sp1} 
       ~%\\
       \cspkw{then}~\cspkw{true}
       \\
       \cspkw{else}~(\begin{array}[t]{l}
                       \cspkw{if}~v1\meta{mID} == \meta{sp_2}
                       ~%\\
                       \cspkw{then}~v2\meta{mID} \notin \{\meta{sp_1}\}
                       \\
                       \cspkw{else}~ \ldots
                       \\
                       \cspkw{else}~v2\meta{mID} == \meta{sp_n}~~)
                     \end{array}
      \end{array}
 \end{array}
$
\\
\midrule
%%%%%%%%%%%%%%%%%%%%%%%%%%%%%%%%%%%%%%%%%%%%%%%%%%%%%%%%%%%%%%%%%%%%%%%%%%%
% rules
%%%%%%%%%%%%%%%%%%%%%%%%%%%%%%%%%%%%%%%%%%%%%%%%%%%%%%%%%%%%%%%%%%%%%%%%%%%
%
%%%%%%%%%%%%%%%%%
% rules: single
%%%%%%%%%%%%%%%%%
$\lsem
       \meta{r} 
 \rsemRS
$
& = &
$\lsem\meta{r}\rsemR$
\\ %
%%%%%%%%%%%%%%%%%
% rules: several
%%%%%%%%%%%%%%%%%
$\lsem
       \meta{r~rS} 
 \rsemRS
$
& = &
$\lsem \meta{r} \rsemR \quad \lsem \meta{rS} \rsemRS$
\\
%%%%%%%%%%%%%%%%%
% rule
%%%%%%%%%%%%%%%%%
$\lsem
       \meta{rID}~\kw{when~} \meta{trig}~\kw{then~}\meta{resp}
 \rsemR
$
& = &
$\meta{rID} = Trigger\meta{rID} \cseq Monitoring\meta{rID} \cseq \meta{rID}$
\\
& & 
$Trigger\meta{rID} = \lsem\meta{trig}, \meta{\alpha_E(resp)},\cspkw{Skip},Trigger\meta{rID}\rsemTG$ 
\\
& & 
$Monitoring\meta{rID} = \lsem\meta{resp,trig,\alpha_E(resp)},Monitoring\meta{rID} \rsemRDS$
\\
\bottomrule
\end{tabular}
\end{table*}

%% file: semantics-triggers.tex
\begin{table*}[t]\small
\caption{\label{table:semantics-triggers}Rules that define a tock-CSP semantics for SLEEC triggers. Additional metavariables used here are as follows:~$\meta{AR}$ for an alphabet~(set) of events, $\meta{sp}$ and $\meta{fp}$ for tock-CSP processes, $\meta{mBE}$ for an $\meta{mBoolExpr}$, and $\meta{MIDs}$ for a list of $\meta{measureID}$ elements.}
\centering
\begin{tabular}{l@{\ } l@{\ } p{.67\textwidth}}
\toprule
% %%%%%%%%%%%%%%%%%%%%%%%%%%%%%%%%%%%%%%%%%%%%%%%%%%%%%%%%%%%%%%%%%%%%%%%%%%%
% % specification

%%%%%%%%%%%%%%%%%%%%%%%%%%%%%%%%%%%%%%%%%%%%%%%%%%%%%%%%%%%%%%%%%%%%%%%%%%%
% trigger
%%%%%%%%%%%%%%%%%%%%%%%%%%%%%%%%%%%%%%%%%%%%%%%%%%%%%%%%%%%%%%%%%%%%%%%%%%%
%
%%%%%%%%%%%%%%%%%
% trigger: event
%%%%%%%%%%%%%%%%%
$\lsem
       \meta{eID,AR,sp,fp} 
 \rsemTG
$
& = &
$\meta{eID} \then \meta{sp} \extchoice \meta{(\Extchoice e: AR @ e \then fp)}$
\\
%%%%%%%%%%%%%%%%%%%%%%%%%%%%%%%%%%%%%%%%
% trigger: event and measure expression
%%%%%%%%%%%%%%%%%%%%%%%%%%%%%%%%%%%%%%%%
$\lsem
       \meta{eID}~\kw{and~} \meta{mBE,AR,sp,fp} 
 \rsemTG
$
& = &
$\begin{array}[t]{l}
   \cspkw{let}~~
   %\\
   %\quad
     MTrigger = 
     \lsem\meta{ \alpha_{ME}(mBE),mBE,sp,fp}\rsemME
     %,0)~~
   \\ 
   \cspkw{within}~~\meta{eID} \then MTrigger \extchoice \meta{(\Extchoice e: AR @ e \then fp)}
 \end{array}  
$
\\
%%%%%%%%%%%%%%%%%%%%%
% measure evaluation
%%%%%%%%%%%%%%%%%%%%%
$\lsem
       \meta{\lseq\rseq,mBE,sp,fp}
 \rsemME
$
& = &
$\cspkw{if~} \meta{norm(mBE)} \cspkw{~then~}\meta{sp}\cspkw{~else~}\meta{fp}$
\\
$\lsem
       \meta{\lseq mID\rseq \cat mIDs ,mBE,sp,fp}
 \rsemME
$
& = &
$0 \startBy (\meta{mID}?v\meta{mID} \then 
  \lsem\meta{ \meta{mIDs},mBE[}v\meta{\meta{mID}/\meta{mID}],sp,fp}\rsemME)$
\\
\bottomrule
\end{tabular}
\end{table*}

%% file: semantics-responses.tex
\begin{table*}[t]\small
\caption{\label{table:semantics-responses}Rules for the tock-CSP semantics of SLEEC responses. Additional metavariables used here are:~$\meta{const}$ for a $\meta{constraint}$, $\meta{ARDS}$ for a set of events, $\meta{mp}$ for a process, $\meta{tU}$ for a $\meta{timeUnit}$, $\meta{n}$ for an index~(a natural number), $\meta{dfts}$ for an element of $\meta{defeaters}$, and $\meta{dft}$ for a $\meta{defeater}$.}
\centering
\begin{tabular}{l@{\ } l@{\ } p{.55\textwidth}}
\toprule
%%%%%%%%%%%%%%%%%%%%%%%%%%%%%%%%%%%%%%%
% response and defeaters: no defeaters
%%%%%%%%%%%%%%%%%%%%%%%%%%%%%%%%%%%%%%%
$\lsem
       \meta{const,trig,ARDS,mp}
 \rsemRDS
$
& = &
$\lsem \meta{const},\meta{trig},\meta{ARDS},\meta{mp}\rsemRP$
\\
%%%%%%%%%%%%%%%%%%%%%%%%%%%%%%%%%%%%%%%%%
% response and defeaters: with defeaters
%%%%%%%%%%%%%%%%%%%%%%%%%%%%%%%%%%%%%%%%%
$\lsem
       \meta{const~dfts,trig,ARDS,mp}
  \rsemRDS
$
& = &
$\begin{array}[t]{l}
   \cspkw{let}~~
   %\\
   %\quad
     \lsem \meta{\lseq const\rseq \cat dfts\filter_{RP},trig,ARDS,mp,1} \rsemLRDS
   \\ 
   \cspkw{within}~~\lsem \meta{\alpha_{ME}(dfts),dfts,\# dfts + 1}\rsemCDS
 \end{array}  
$
\\
\midrule

\begin{comment}
exp ::=  value | exp "plus" exp | exp "minus" exp

[[v]]EXP =  v

[[exp1 "plus" exp2]]EXP =  [[exp1]]EXP \textbf{+} [[exp2]]EXP
\end{comment}

%%%%%%%%%%%%%%%%%%%%%%%%%%%%%%%%%%%%%%%%%%%%%%%%%%%%%%%%%%%%%%%%%%%%%%%%%%%
% response
%%%%%%%%%%%%%%%%%%%%%%%%%%%%%%%%%%%%%%%%%%%%%%%%%%%%%%%%%%%%%%%%%%%%%%%%%%%
%
%%%%%%%%%%%%%%%%%%
% response: event
%%%%%%%%%%%%%%%%%%
$\lsem
       \meta{eID},\meta{trig},\meta{ARDS},\meta{mp}
 \rsemRP
$
& = &
$\meta{eID} \then \cspkw{Skip}$
\\
%%%%%%%%%%%%%%%%%%%%%%%%%
% response: event within
%%%%%%%%%%%%%%%%%%%%%%%%%
$\lsem
       \meta{eID}~\kw{within}~\meta{v}~\meta{tU},\meta{trig},\meta{ARDS},\meta{mp}
 \rsemRP
$
& = &
$\meta{norm(v,tU)} \startBy (\meta{eID} \then \cspkw{Skip})$
\\
%%%%%%%%%%%%%%%%%%%%%%%%%%%%%%%%%%%
% response: event within otherwise
%%%%%%%%%%%%%%%%%%%%%%%%%%%%%%%%%%%
$\lsem
  \meta{eID}~\kw{within}~\meta{v}~\meta{tU}~\kw{otherwise}~\meta{resp},\meta{trig},\meta{ARDS},\meta{mp}
 \rsemRP
$
& = &
$(\meta{eID} \then \cspkw{Skip}) \interrupt_{\meta{norm(v,tU)}} (\lsem\meta{ resp,trig,ARDS,mp}\rsemRDS)$
\\
%%%%%%%%%%%%%%%%%%%%%%%%%%%%%
% response: not event within
%%%%%%%%%%%%%%%%%%%%%%%%%%%%%
$\lsem
       \kw{not}~\meta{eID}~\kw{within}~\meta{v}~\meta{tU},\meta{trig},\meta{ARDS},\meta{mp}
 \rsemRP
$
& = &
$\cspkw{Wait}(\meta{norm(v,tU)})$
\\
\midrule
%
%%%%%%%%%%%%%%%%%%%%%%%%%%%%%%%%%%%%%%%%%%%%%%%%%%%%%%%%%%%%%%%%%%%%%%%%%%%
% local monitoring definitions
%%%%%%%%%%%%%%%%%%%%%%%%%%%%%%%%%%%%%%%%%%%%%%%%%%%%%%%%%%%%%%%%%%%%%%%%%%%
$\lsem
   \meta{\lseq resp\rseq,trig,AR,mp,n}
 \rsemLRDS
$
& = &
$Monitoring\meta{n} = \lsem \meta{resp},\meta{trig},\meta{AR},\meta{mp} \rsemRDS$,~provided~$\meta{resp} \neq \cspkw{NoRep}$
\\
$\lsem
   \meta{\lseq \cspkw{NoRep}\rseq,trig,AR,mp,n}
 \rsemLRDS
$
& = &
$Monitoring\meta{n} = \begin{array}[t]{l}
                           \lsem \meta{trig,AR,mp,}Monitoring\meta{n} \rsemTG
                           \\
                           \extchoice 
                           \\
                           (\Extchoice e: \meta{AR} @ e \then Monitoring\meta{n})
                         \end{array}$
\\
$\lsem
   \meta{\lseq resp\rseq\cat resps,trig,AR,mp,n}
 \rsemLRDS
$
& = &
$\begin{array}[t]{l}
  \lsem \meta{\lseq resp\rseq,trig,AR,mp,n} \rsemLRDS
  \\
  \lsem \meta{resps,trig,AR,mp,n+1} \rsemLRDS
 \end{array}
$
\\
\midrule
%
%%%%%%%%%%%%%%%%%%%%%%%%%%%%%%%%%%%%%%%%%%%%%%%%%%%%%%%%%%%%%%%%%%%%%%%%%%%
% check and execute defeaters
%%%%%%%%%%%%%%%%%%%%%%%%%%%%%%%%%%%%%%%%%%%%%%%%%%%%%%%%%%%%%%%%%%%%%%%%%%%
%
%%%%%%%%%%%%%%%%%%
% check defeaters
%%%%%%%%%%%%%%%%%%
$\lsem
       \meta{\lseq\rseq,dfts,n}
 \rsemCDS
$
& = &
$\lsem\meta{dfts,}Monitoring1\meta{,n}\rsemEDS$
\\
$\lsem
       \meta{\lseq mID\rseq \cat mIDs,dfts,n}
 \rsemCDS
$
& = &
$0 \startBy (\meta{mID}?v\meta{mID} \then \lsem\meta{mIDs},\meta{dfts[}v\meta{\meta{mID}/\meta{mID}],n}\rsemCDS)$
\also
%
%\\
%%%%%%%%%%%%%%%%%%%%
% execute defeaters
%%%%%%%%%%%%%%%%%%%%
%%%%%%%%%%%%%%%%%%%%%%%%%%%%%%%
% defeater: unless no response
%%%%%%%%%%%%%%%%%%%%%%%%%%%%%%%
$\lsem
       \kw{unless}~\meta{mBE,fp,n}
 \rsemEDS
$
& = &
$\cspkw{if~} \meta{norm(mBE)} \cspkw{~then~}Monitoring\meta{n~else~ fp}$
\\
%%%%%%%%%%%%%%%%%%%%%%%%%%%%%%%%%
% defeater: unless with response
%%%%%%%%%%%%%%%%%%%%%%%%%%%%%%%%%
$\lsem
       \kw{unless}~\meta{mBE} \kw{~then~} \meta{resp,fp,n}
 \rsemEDS
$
& = &
$\cspkw{if~} \meta{norm(mBE)} \cspkw{~then~}Monitoring\meta{n~else~ fp}$
\\
%%%%%%%%%%%%%%%%%%%%%%%%%%%%%%%%%
% defeaters
%%%%%%%%%%%%%%%%%%%%%%%%%%%%%%%%%
$\lsem
       \meta{dfts~dft,fp,n}
 \rsemEDS
$
& = &
$\lsem\meta{ dft,\lsem dfts,fp,n-1 \rsemEDS,n} \rsemEDS$
\\
\bottomrule
\end{tabular}
\end{table*}

%% file: vandv.tex
%%%%%%%%%%%%%%%%%%%%%%%%%%%%%%%%%%%%%%%%%%%%%%%%%%%%%%%%%%%%
%%%%%%%%%%%%%%%%%%%%%%%%%%%%%%%%%%%%%%%%%%%%%%%%%%%%%%%%%%%%

When writing SLEEC rules, it is possible to make a mistake and introduce redundant or conflicting rules, especially given the possibility that these rules are provided by stakeholders with different expertise (lawyers, ethicists, sociologists, etc.) and comprise complex defeaters. Redundant rules  may help stakeholders to understand the consequences of the rules; for verification, however, these rules are unnecessary and so should be flagged.  Conflicting rules, on the other hand, mean that there is no implementation that can satisfy them all.  They need to be flagged and the conflict needs to be resolved.  In Section~\ref{subsec:conflict}, we present an approach that uses the semantics of our rules presented above to detect conflicts, and in Section~\ref{subsec:verification}, we present redundancy checks.  Finally, in Section~\ref{subsec:verification}, we discuss the verification of an agent model against a set of SLEEC rules.
%%%%%%%%%%%%%%%%%%%%%%%%%%%%%%%%%%%%%%%%%%%%%%%%%%%%%%%%%%%%
\subsection{SLEEC conflict detection}
\label{subsec:conflict}
%%%%%%%%%%%%%%%%%%%%%%%%%%%%%%%%%%%%%%%%%%%%%%%%%%%%%%%%%%%%
Two rules $\meta{r1}$ and $\meta{r2}$ are conflict free if there is no scenario in which both rules apply and the restriction of $\meta{r1}$ makes it not possible to satisfy the restriction of $\meta{r2}$, or vice-versa. Conjunction is specified in CSP using parallelism.  So,  roughly speaking, %formally 
conflict freedom requires the process formed by the parallel %ism 
combination of the processes for $\meta{r1}$ and $\meta{r2}$ never to reach a state in which the only event that can happen, if any, is $tock$. In this case, a system that satisfies both rules cannot make useful progress.    

In practical terms, we only need to check for conflict between rules that have an overlap in their alphabet of events. If the rules have no such overlap, the restrictions they impose cannot interfere with each other. Moreover, overlap in the alphabet of measures is irrelevant, as rules do not need to agree on the reading of measures.  The measures represent information about the system and the environment that is available at any time. 

Table~\ref{table:conflict-sem} presents the function $\meta{\lsem r1,r2\rsemCP}$, which defines the conjunction process $\meta{idCC(r1,r2)}$ for the rules $\meta{r1}$ and $\meta{r2}$. %The definition of $\meta{idCC(r1,r2)}$ uses clauses \cspkw{let} and \cspkw{within}.  
In the \cspkw{within} clause of this definition, we compose the processes $\meta{id(r1)}$ and $\meta{id(r2)}$ (which define the semantics of $\meta{r1}$ and $\meta{r2}$) in parallel~($\lpar... \rpar$), synchronising on their common alphabet of events, i.e., on the intersection alphabets~$\meta{\alpha_E(r1)} \cap \meta{\alpha_E(r2)}$ of their alphabets.

An additional parallel process $Env$ captures the environment in which the rules are considered, by recording the values of the measures for sharing between $\meta{id(r1)}$ and $\meta{id(r2)}$.
The definition of $Env$ is given in the \cspkw{let} clause as the interleaving, that is, the parallel combination without synchronisation, of processes $Env\meta{e}$ for each event $\meta{e}$ in the alphabet of measures $\meta{\alpha_M(r1,r2)}$ of $\meta{r1}$ and $\meta{r2}$.  These processes input the value of the measure $\meta{e}$ when a rule first requires that measure~($\meta{e}?x)$.  They then record the value $x$ input as a parameter for another process $VEnv\meta{e}$, which outputs $x$~($\meta{e!x}$) whenever a rule needs that measure.  With $Env$ we ensure that, for conflict checking, the rules are considered when the measures take the same value.  

Using $\meta{\lsem r1,r2\rsemCP}$, we define conflict freedom for the rules $\meta{r1}$ and $\meta{r2}$ below. For that, we use the process operator `$P \after t$', which defines the process that behaves like $P$ after it has already engaged in its trace of events $t$.

\begin{definition}
    The rules $\meta{r1}$ and $\meta{r2}$ are conflict free, if, and only if, for every trace $t_1$ of $\meta{\lsem r1,r2\rsemCP}$, there is a trace $t_2$ of $\meta{\lsem r1,r2\rsemCP} \after t_1$ that contains at least one $tock$ and at least one event different from $tock$.
    \label{definition:conflict-freedom}
\end{definition}

\begin{table}[]
\caption{Conjunction of rules $\meta{r1}$ and $\meta{r2}$} \label{table:conflict-sem}
\begin{tabular}{l@{\ } l@{\ 
} p{62mm}}
\toprule
\also
$\lsem 
    \meta{r1,r2}
  \rsemCP
$
& = &
$\begin{array}[t]{l}
    \meta{idCC(r1,r2)} = \cspkw{let}
    \\
    \
      \begin{array}{l}
    Env = \meta{\Interleave~e: \alpha_M(r1,r2) @} Env\meta{e}
    \\
    Env\meta{e} = \meta{e}?x \then VEnv\meta{e}(x)
    \\
    VEnv\meta{e}(x) = \meta{e}!x \then  VEnv\meta{e}(x)
    \\
    \cspkw{within}
    \\
    \quad
      \begin{array}[t]{l}
         (\meta{id(r1)}
         %\\
         %\quad
           \lpar \meta{\alpha_E(r1)} \cap \meta{\alpha_E(r2)} \rpar
        %\\
        \meta{id(r2)})
        \\
        \quad
          \lpar \meta{\alpha_M(r1,r2)} \rpar 
        \\
        Env
    \end{array}
  \end{array}
  \end{array}
$
\\
\bottomrule
\end{tabular}
\end{table}

\noindent%
With this definition, we require that, at no point, enforcing both rules, as defined by the process $\meta{\lsem r1,r2\rsemCP}$, leads to a deadlock, so that no more events are possible, or to a situation in which there is no deadlock, but only the passage of time can be observed.  The latter scenario is a timed deadlock:~time can progress, but no event is possible.  

Definition~\ref{definition:conflict-freedom} is given in terms of the semantics, that is, the set of traces, of the conjunction process $\meta{\lsem r1,r2\rsemCP}$.  For automation, we can check conflict freedom using FDR using two assertions. The first is a standard FDR assertion for deadlock freedom, and the second is an assertion based on our mechanisation of a timed-deadlock freedom check in the context of tock-CSP that is inspired by work in~\cite{Roscoe2013a}. 

\begin{example}
    Listing~\ref{lst:deadlock-conflict} presents another rule (\textsf{RuleA}) for the firefighter UAV. This rule requires that, if the battery reaches a critical level, and there is no risk of fire nearby, %and so no risk to life, 
    as indicated by the \textsf{temperature} measure, then the robot should return to base so that it can continue to work at a later point.  We can imagine that, if there is risk of fire, the UAV should continue its mission even if it means that it will exhaust its battery in action.  However, \textsf{RuleA} is in conflict with \textsf{Rule3} from Listing~\ref{lst:trigger}. We show in Figure~\ref{fig:RuleARule3} the conjunction CSP process for the two rules.  In this case, both rules restrict the \textsf{GoHome} event and use just one measure, \textsf{temperature}. So, the $Env$ process is just the $Envtemperature$ process for this measure.  The deadlock check (using the FDR model checker) gives a counterexample that indicates the reason for the deadlock. Namely, it provides a trace with the events $BatteryCritical$ and $temperature.20$, and after 13 occurrences of $tock$, then the event $SoundAlarm$, followed by 47 occurrences of $tock$. In this case, we are identifying a time unit with 1s. So, the counterexample, indicates that if \textsf{RuleA} is triggered, and after 13s, \textsf{Rule3} is triggered, then, after another 47s, we have a deadlock, as \textsf{RuleA} requires $GoHome$ to take place, but \textsf{Rule3} forbids it. \hfill $\Box$
\end{example}

\begin{lstlisting}[language = SLEEC, label={lst:deadlock-conflict}, caption={Conflicting rule for a firefighter robot},captionpos=b, float=tp, frame = single]
rule_start
  RuleA when BatteryCritical and temperature < 25 
        then GoHome within 1 minute 
rule_end
\end{lstlisting}

\begin{figure*}
    \begin{displaymath}
        RuleARule3 = \begin{array}[t]{l}\cspkw{let}
        %\\
        %\quad
          \begin{array}[t]{l}
		  Envtemperature = temperature?x \then VEnvtemperature(x)
            \\
            VEnvtemperature(x) = temperature!x \then VEnvtemperature(x)
            \\
            Env = Envtemperature
          \end{array}
        \\
        \cspkw{within}
        %\\
        %\quad
          (RuleA \lpar \{GoHome\} \rpar Rule3) \lpar \lchan temperature\rchan \rpar	Env
      \end{array}    
    \end{displaymath}
  \caption{\label{fig:RuleARule3} Conjunction process for \textsf{RuleA} in Listing~\ref{lst:deadlock-conflict} and \textsf{Rule3} in Listing~\ref{lst:trigger}. }
\end{figure*}

\noindent
If the assertion for deadlock freedom holds, there is no guarantee that timed deadlock freedom holds.  

\begin{example} 
Listing~\ref{lst:example-conflict} presents two other conflicting rules for the firefighter UAV. %A fragment code of corresponding CSP processes are shown in Figure \ref{fig:example-conflict-CSP} where lines 1-8 describe the \textsf{RuleA} process having two measures \textsf{temperature} and \textsf{personNearBy}). Process \textsf{RuleB} is defined similarly based on the semantics. Lines between 13-17 demonstrates the conflict checking process and the \textsf{deadlock} assertion. 
In the case of these rules, their conjunction does not lead to a deadlock.  It is the case, however, that there is a situation in which the only possible behaviour allowed by these two rules is the passage of time. The check for timed deadlock freedom provides the following counterexample. First $BatteryCritical$ happens, so that both rules are triggered. Afterwards, the measures $personNearby$ and $temperature$ are read and the values provided are $true$ and $33$.  So, \textsf{RuleC} requires the robot to $GoHome$ and \textsf{RuleD} requires it to $SoundAlarm$ instead.  We do not have a deadlock, as time can pass in the absence of a deadline.  It so happens, however, that \textsf{RuleC} forbids $SoundAlarm$ and \textsf{RuleD} forbids $GoHome$ (because the two events are in the alphabets of both rules, and $SoundAlarm$ is not mentioned in the relevant defeater of \textsf{RuleC}, while $GoHome$ is not mentioned in the relevant defeater of \textsf{RuleD}). So, neither event can happen. \hfill $\Box$
\end{example}

\begin{lstlisting}[language = SLEEC, label={lst:example-conflict}, caption={Conflicting rule for the firefighter UAV},captionpos=b, float=tp, frame = single]
rule_start
  RuleC when BatteryCritical 
        then CameraStart
        unless personNearby then GoHome 
        unless temperature > 35 then SoundAlarm

  RuleD when BatteryCritical
        then CameraStart
        unless personNearby then SoundAlarm 
        unless temperature > 35 then GoHome
rule_end
\end{lstlisting}

\noindent
If a pair of rules are not conflicting, but their alphabets overlap, then one of them may be redundant.  We next consider how to check for redundancy.

%%%%%%%%%%%%%%%%%%%%%%%%%%%%%%%%%%%%%%%%%%%%%%%%%%%%%%%%%%%%
\subsection{Detection of superfluous rules}
\label{subsec:redundancy}
%%%%%%%%%%%%%%%%%%%%%%%%%%%%%%%%%%%%%%%%%%%%%%%%%%%%%%%%%%%%

For a pair of rules $\meta{r1}$ and $\meta{r2}$ that have overlapping alphabets and are not conflicting, %$\lsem \meta{r1,r2} \rsemUC$ 
we define redundancy below, using $t \filter E$ to denote the trace obtained from $t$ by removing all events that are not in the set $E$.  
\begin{definition}
    For conflict-free rules $\meta{r1}$ and $\meta{r2}$, we say $\meta{r2}$ is redundant with respect to $\meta{r1}$ if, and only if, for every trace $t_1$ of $\meta{id(r1)}$, there is a trace of $t_2$ of $\meta{\lsem r1,r2\rsemCP}$, such that, $t_1 \filter \meta{\alpha_E(r_1)} = t_2 \filter \meta{\alpha_E(r_1,r_2)}$.
    \label{definition:redundancy}
\end{definition}
\noindent
First of all, we observe that the traces of the process that characterises a rule identify the behaviours allowed by the rule.  So, the smaller that set of traces, the more restrictive is that rule.  In this context, however, the reading of measures is irrelevant, since, as already said, rules use measures just to obtain information that indicates how the events are to restricted. So, in Definition~\ref{definition:redundancy}, we characterise a rule $r2$ as redundant, with respect to another rule $\meta{r1}$, by considering the traces $t_1 \filter \alpha_E(r_1)$, where $t_1$ is a trace of $\meta{r1}$ and $\alpha_E(r_1)$ is the set of events of $\meta{r1}$ or, more precisely, the set of CSP events that represent the SLEEC events of $\meta{r1}$.  These traces characterise the restrictions of $\meta{r1}$.  Similarly, the traces $t_2 \filter \alpha_E(r_1,r_2)$ characterise the restrictions of $\meta{r1}$ and $\meta{r2}$.  If every behaviour allowed by $\meta{r1}$ is also allowed by $\meta{r1}$ and $\meta{r2}$, then $\meta{r2}$ imposes no additional restrictions, and it is, therefore, redundant. 
 
The mechanisation of this check is direct, since trace inclusion in CSP corresponds to refinement, and ignoring events can be captured using the hiding CSP operator ($\filter$).  

\begin{example}
 In Listing \ref{lst:trigger}, \textsf{Rule1} is weaker, as it does not have a deadline, and can be eliminated. This can be automatically checked using the FDR model checker via a trace refinement.  Since the refinement holds, there is no counterexample, but a clear indication of the weaker rule between the two.  \hfill $\Box$
\end{example}
\noindent
Normally, a rule $\meta{r2}$ should not be redundant with respect to another rule $\meta{r1}$ if $\meta{r2}$ involves events not referenced in $\meta{r1}$.  This is, however, not necessarily the case, since the responses that refer to the extra events may be unreachable.  So, in general, it is worth checking every pair of non-conflicting rules with overlapping alphabets of events.  It is also possible to check for unreachable responses. Next, however, we consider autonomous agent conformance to SLEEC rules. 

Our experience with the case studies presented in this paper and with a number of other examples of autonomous agents, 
%(eight, beyond those presented here) and 
as well as discussions with SLEEC experts suggest that the number of SLEEC rules for an autonomous agent will not run into the hundreds:~it is more like tens, if that. Moreover, a single rule is unlikely to have a very long or very deep list of defeaters. So, although our checks require a pairwise analysis of the rules, we expect that the checks for conflicts and redundancy within a SLEEC specification for an autonomous agent will remain tractable.  Importantly, as we avoid dealing with the whole set of rules in a single check, model checking is also likely to remain feasible. The treatment of more complex data types provided by measures, however, are likely to impose a challenge. 
%%%%%%%%%%%%%%%%%%%%%%%%%%%%%%%%%%%%%%%%%%%%%%%%%%%%%%%%%%%%
\subsection{Verification of compliance with SLEEC rules} \label{subsec:verification}
%%%%%%%%%%%%%%%%%%%%%%%%%%%%%%%%%%%%%%%%%%%%%%%%%%%%%%%%%%%%

\begin{figure*}
\centering
  \includegraphics[scale=.73]{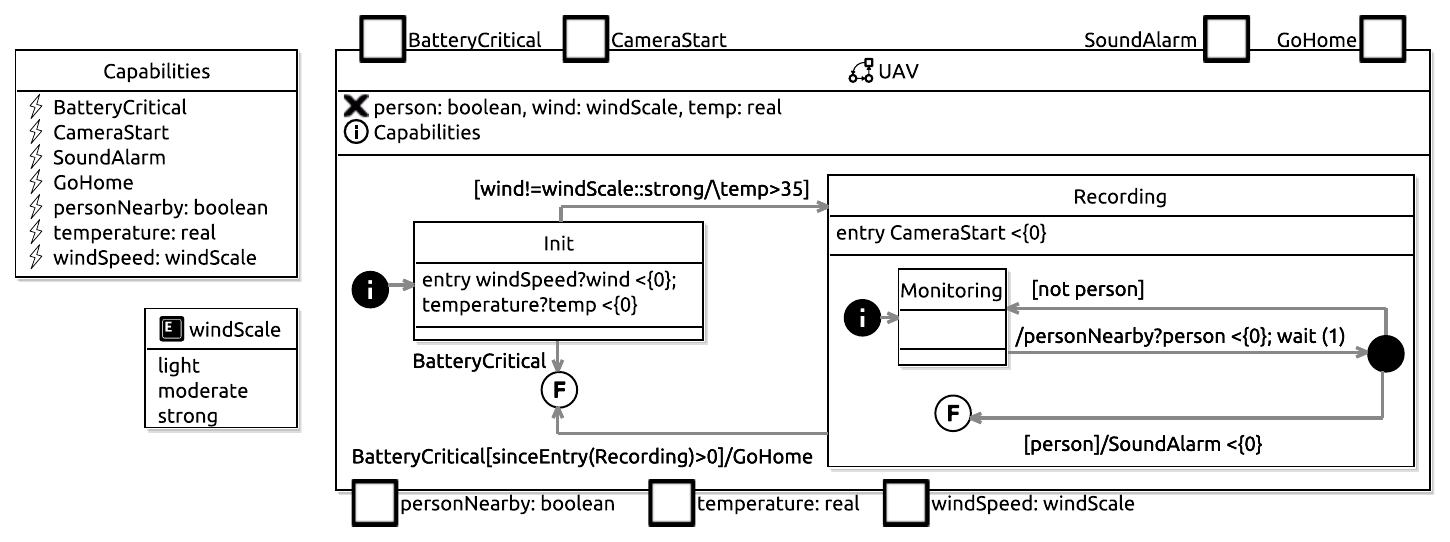}
  \caption{\label{figure:simple-firefighter-robochart} Sketch of RoboChart model for a simple firefighter UAV}
\end{figure*}

{\setlength{\fboxsep}{0pt}\setlength{\fboxrule}{1pt}
\begin{figure*}[t]
\centering
\fbox{
\includegraphics[width=16cm]%,height=10cm]
{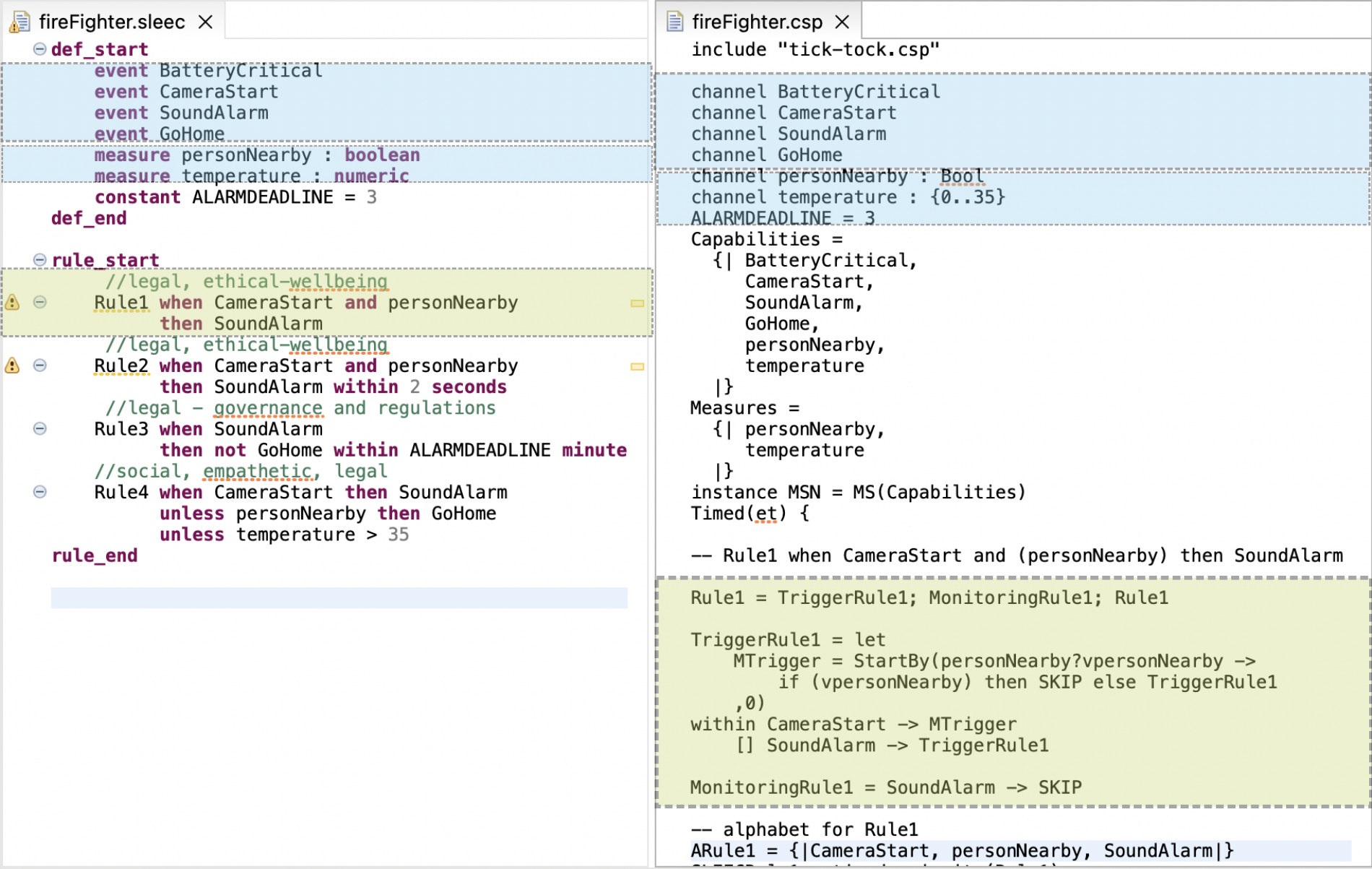}}
%\vspace{-1em}
  \caption{Tool for editing and supporting reasoning about SLEEC rules}\label{fig:tooling}

  \vspace*{-3mm}
\end{figure*}
}

This section describes our method for checking a system under verification (SUV) against a SLEEC rule $\meta{r}$ by means of refinement in tock-CSP. To that end, we assume the existence of a tock-CSP model for the SUV. %, but recall that it is possible to automatically generate such models. 
Such models can be generated automatically from other design models, or can be devised manually by system developers with CSP modelling expertise. Here, we consider examples where a RoboChart~\cite{Robochart} modelfor the agent is available and used as a basis to generate a tock-CSP model $SUV$ automatically.  

The notion of conformance $\meta{r} \models_{TT} SUV$ that we adopt is defined below. In words, it corresponds to traces refinement in tock-CSP, where the specification is defined in terms of the process $\lsem \meta{r} \rsemR$ that captures the semantics of $\meta{r}$ (cf.~Table~\ref{table:semantics}). Traces refinement in tock-CSP ensures that the events of the SUV occur in the order and time specified, so that time budgets and deadlines are respected. Like in the check for redundancy, refinement disregards the measures; here, however, we require the values of the measures recorded in the specification and in the SUV to be the same.  (In the definition of redundancy from Section~\ref{subsec:redundancy}, this is ensured by the conjunction process.)

\begin{definition}
  An SUV conforms to a rule \textsf{r}, written $\meta{r} \models_{TT} SUV$, where $SUV$ is the tock-CSP model of SUV, if, and only if, for every trace $t_1$ of the process $SUV; Stop$, there is a trace $t_2$ of $\lsem \meta{r} \rsemR$ such that: (1)~$t_1 \filter \meta{\alpha_E(r)} = t_2 \filter \meta{\alpha_E(r)}$ and; (2)~for every event $e$ of $\meta{\alpha_E(r)}$ in a position $i$ of these traces, for every measure $m$ in $\meta{\alpha_M(r)}$, the value of $m$ recorded in $t_1$ and $t_2$ at position $i$ are the same.  
  \label{definition:conformance}
\end{definition}

\noindent
We consider the process $SUV; Stop$, rather than just $SUV$, because the processes that give semantics to a rule do not terminate.  If SUV terminates, subsequent composition with $Stop$ ensures that we do not erroneously flag a problem just because the rule does not allow termination.  

According to Definition~\ref{definition:conformance}, a conforming SUV may engage in additional events and read additional measures.  The value $v$ of a measure $m$ at $i$ in a trace $t$ is that in the last event $m.v$ before the occurrence of the $i$-th event of $\meta{\alpha_E(r)}$ in $t$. Conformance requires that, if a rule reads a measure, a conforming SUV must read that measure as well.  Moreover, when checking conformance, we consider traces based on the same values for those measures. 
 
As mentioned earlier, the mechanisation of conformance checking is based on refinement, but the specification is a weakening of $\lsem \meta{r} \rsemR$, with respect to refinement, to allow occurrence of additional events and any order in the reading of measures.  

We illustrate the verification process using a simplified model of the control software of the firefighting UAV. In the next section, we consider additional examples.   

\begin{example}
    As said, for modelling we use RoboChart. In Figure~\ref{figure:simple-firefighter-robochart}, we show a RoboChart interface \RC{Capabilities} that declares those capabilities of the firefighter UAV that we identified in the SLEEC specification.  Other interfaces in the model may declare additional capabilities not related to SLEEC concerns, but needed to implement the firefighter UAV mission.  We also show a RoboChart state machine called \RC{UAV} that specifies the control software of our simple firefighter in terms of these \RC{Capabilities} and using local variables~(\RC{person}, \RC{wind}, and \RC{temp}).  
    In the initial state \RC{Init} of \RC{UAV}~(the target of the transition out of the initial junction indicated by a dark circle with an \RC{i}), an \RC{entry} action reads the \RC{windSpeed}, recording it in the local variable \RC{wind}, and the \RC{temperature}, recording it in \RC{temp}.  The notation `\RC{$<$\{0\}}' specifies that these inputs need to be immediately available. There are two transitions out of \RC{Init}.  The first has the event \RC{BatteryCritical} as a trigger. If this event happens, the \RC{UAV} cannot proceed, and terminates by transitioning to the final state, indicated by a clear circle with an \RC{F}.  The other transition has no trigger, but a guard that requires the wind not to be \RC{strong}~(\RC{wind != windScale::strong}, where \RC{windScale} is the enumeration type of \RC{wind} defined on the left in Figure~\ref{figure:simple-firefighter-robochart}), and the temperature to be high~(\RC{temp $>$ 35}), indicating a possible fire.  That transition leads to a composite state \RC{Recording} whose \RC{entry} action starts the camera.  Its own state machine is concerned with whether there is a \RC{personNearby}.  Every 1s, this state machine reads that measure and records it in the variable \RC{person}. Depending on whether there is such a person or not, it raises the event \RC{SoundAlarm}.  A transition out of \RC{Recording} ensures that, when the \RC{BatteryCritical} is signalled, the UAV goes back to base by raising the event \textsf{GoHome}. The guard ensures that the amount of time since the state \RC{Recording} has been entered~(\RC{sinceEntry(Recording)}) is greater than 0, so that the check for the presence of a person is carried out before returning. 
    
    There are several simplifications in this example, but our focus is on the rules in Listing~\ref{lst:trigger}.  We have identified that \textsf{Rule1} is redundant, so we do not need to be concerned with it.  Our technique identifies that the model satisfies \textsf{Rule2}, but not \textsf{Rule3}. The counterexample provided by the FDR model checker has the following events:
    \begin{displaymath}
        windSpeed.light, temperature.36,
        \\
        CameraStart, personNearby.true,tock
        \\
        SoundAlarm, BatteryCritical, GoHome
    \end{displaymath}
    This counterexample is a trace that leads to a forbidden event, here $GoHome$.  The trace corresponds to a scenario in which, in the \RC{Init} state, the measures \RC{windSpeed} and \RC{temperature} read are as $light$ and $36$, respectively.  With that, in the state \RC{Recording}, the camera is started, when there is a \RC{personNearby}. So, after 1~s~(i.e., one $tock$), the alarm is sounded, but the battery is indicated as critical.  In this situation the \RC{UAV} goes home, but \textsf{Rule3} forbids that for 5 minutes. Indeed, the projection of this \RC{UAV} trace to the events of \textsc{Rule3} is $SoundAlarm, GoHome$, which is not a trace of the process for \textsf{Rule3} (without the events that refer to measures). So, considering Definition~\ref{definition:conformance}, condition~(1) is not satisfied. \hfill $\Box$
\end{example}

\noindent%
Before providing additional examples in the next section, we note that the relatively low complexity of SLEEC rules is expected to make the verification of SUV compliance with each individual rule feasible---under the assumption that the SUV tock-CSP model is itself of manageable size. As is often the case with model checking, this assumption may not always hold because of state explosion, in particular as the FDR model checker is not optimised for dealing with timed (i.e., tock-CSP) models despite support them. On the positive side, RoboChart is part of a framework that includes support for alternative verification approaches, based on theorem proving, simulation, and testing~\cite{CBBCFMRS21}. In particular, theorem proving is promising, and amenable to automation if we use automatically generated semantics, like that of SLEEC. 

%% file: evaluation.tex
%%%%%%%%%%%%%%%%%%%%%%%%%%%%%%%%%%%%%%%%%%%%%%%%%%%%%%%%%%%%
%%%%%%%%%%%%%%%%%%%%%%%%%%%%%%%%%%%%%%%%%%%%%%%%%%%%%%%%%%%%
\section{Evaluation: Tool support and additional \newline case study}
\label{sec:evaluation}
%%%%%%%%%%%%%%%%%%%%%%%%%%%%%%%%%%%%%%%%%%%%%%%%%%%%%%%%%%%%
%%%%%%%%%%%%%%%%%%%%%%%%%%%%%%%%%%%%%%%%%%%%%%%%%%%%%%%%%%%%

In this section, we present our efforts to validate our work beyond the firefighter UAV case study presented as a running example in the previous sections.  In Section~\ref{section:tool}, we present a mechanisation of the semantics in Section~\ref{sec:semantics} to support editing of SLEEC rules and automatic generation of tock-CSP scripts. The close relationship between the definition of the semantics and its mechanisation provides validation for the work via evidence that there are enough definitions, and that they produce valid tock-CSP processes. Section~\ref{section:rad} presents another case study, namely, an assistive dressing robot.  The SLEEC rules for this example have been developed in collaboration with SLEEC experts, and we have used our tool to validate the rules.  In addition, we have verified a design of that robot with respect to our rules. 

%%%%%%%%%%%%%%%%%%%%%%%%%%%%%%%%%%%%%%%%%%%%%%%%%%%%%%%%%%%%
\subsection{SLEEC tool}
\label{section:tool}
%%%%%%%%%%%%%%%%%%%%%%%%%%%%%%%%%%%%%%%%%%%%%%%%%%%%%%%%%%%%

We have implemented the SLEEC syntax~(Figure \ref{lst:BNF}) in Eclipse with approximately 120~lines of Xtext~\cite{XText} code.  
The translation of SLEEC documents to tock-CSP is based on the definitions presented in Tables~\ref{table:semantics}--\ref{table:semantics-responses} and is implemented in Xtend (approximately 700 lines of code)~\cite{XTend}, a lightweight version of Java.

The implementation of $\meta{norm(mBE)}$ allows seconds, minutes, hours and days in the concrete syntax and it normalises each value to seconds in the current implementation. The resulting tool is described in~\cite{YBJCC23}.
We tested a wide range of different rule structures from Tables~~\ref{table:semantics}--\ref{table:semantics-responses} specified in our SLEEC language.
All code and the models are publicly available~\cite{sleec}.

For the semantics, we translate from the trace-based definitions of conflict, redundancy, and conformance, to refinement checks via a mechanisation of $tock$-CSP~\cite{BaxterRC22} for verification using the CSP model-checker FDR~\cite{FDR}. This enables the automatic analysis of SLEEC rules and verification of conformance against system models with $tock$-CSP semantics, such as in the case of RoboChart models. 

Figure~\ref{fig:tooling} shows a screenshot of our tool, where we can see the encoding of the SLEEC definitions and rules from Listings~\ref{lst:defBlock} and \ref{lst:trigger} in the left pane, and the automatically generated tock-CSP script for the SLEEC specification in the right pane. We note that, because model checking operates with finite models, measures of the type $Int$ need to be specified using finite intervals such as \{0..35\}.   

%%%%%%%%%%%%%%%%%%%%%%%%%%%%%%%%%%%%%%%%%%%%%%%%%%%%%%%%%%%%
%%%%%%%%%%%%%%%%%%%%%%%%%%%%%%%%%%%%%%%%%%%%%%%%%%%%%%%%%%%%

%%%%%%%%%%%%%%%%%%%%%%%%%%%%%%%%%%%%%%%%%%%%%%%%%%%%%%%%%%%%
\subsection{Robot Assistive Application}
\label{section:rad}
%%%%%%%%%%%%%%%%%%%%%%%%%%%%%%%%%%%%%%%%%%%%%%%%%%%%%%%%%%%%
We present here our second case study, a robotic assistive dressing~(RAD) system tasked with aiding a physically impaired user with dressing adapted from the solution presented in~\cite{camilleri2022DCS}. A secondary function of RAD is to monitor the health of the user, who is liable to fall. When falls are detected, RAD is expected to contact support services.  Health and fall monitoring is achieved through a smart watch worn by the user and by visual sensors mounted on the platform and in the user's home. To communicate with the user, RAD is equipped with voice recognition and speech modules. RAD can communicate with a support operator located off site by transmitting audio and video feeds. Finally, RAD can control temperature, lighting and the opening/closing of the room curtains through home automation functionality.

\begin{figure*}[t]
\begin{center}
    \centering
    \begin{subfigure}[m]{0.5\textwidth}
        \centering
        \includegraphics[width=\textwidth]{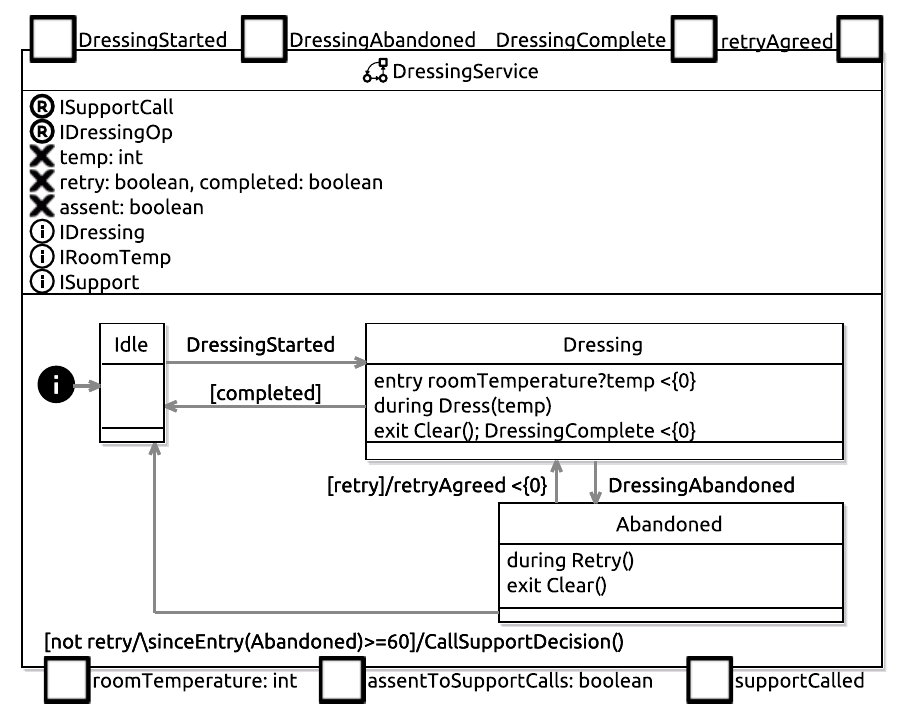}
        \caption{DressingService state machine}
    \end{subfigure}%
    ~ 
    \begin{subfigure}[m]{0.5\textwidth}
        \centering
         \includegraphics[width=\textwidth]{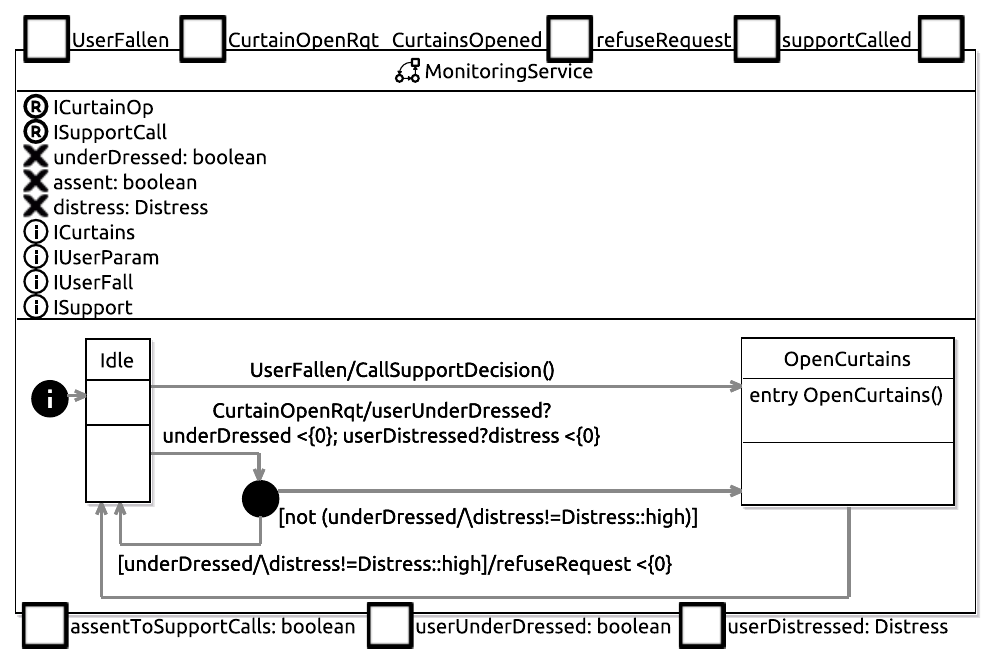}
        \caption{MonitoringService state machine}
    \end{subfigure}
 %\vspace{-0.5em}
  \caption{RoboChart state machines of the RAD application}\label{fig:RAD}
  \end{center}
  %\vspace*{-4mm}
\end{figure*}

RAD's control software is specified as a RoboChart model that includes multiple state machines defining parallel behaviour. The two state machines relevant to the discussion here are \RC{DressingService} and \RC{MonitoringService}; they are depicted in Figure~\ref{fig:RAD}. \RC{DressingService} specifies the software for dressing of users, and interacts with the platform via events such as \RC{DressingStarted} and \RC{DressingAbandoned}. \RC{MonitoringService} mediates the opening of a room's curtains, and can call support if needed. In what follows, we describe each state machine in further detail.

In \RC{DressingService} the state machine is initially in an \RC{Idle} state. A transition to a \RC{Dressing} state is triggered by the event \RC{DressingStarted}, corresponding to a request from the user. In that state's \RC{entry} action the current \RC{roomTemperature} is input into a local variable \RC{temp}, and then the behaviour is given by a call to an operation \RC{Dress} with the current temperature passed as a parameter. Here, \RC{Dress} is a software operation that captures the time the actual dressing can take, after which it sets the value of a Boolean variable \RC{completed} to true. Because \RC{Dress} is called in a \RC{during} action, this behaviour can be interrupted by any of \RC{Dressing}'s outgoing transitions:~either because \RC{completed} is true, or as a result of \RC{DressingAbandoned} being triggered. In both cases, before \RC{Dressing} is exited, its \RC{exit} action is executed:~it calls an operation \RC{Clear} that sets the variables \RC{completed} and \RC{retry} to false followed by an output on \RC{DressingComplete} to indicate that dressing has completed, irrespective of whether it has succeeded. In \RC{Abandoned} there is a call to an operation \RC{Retry}. This operation captures a protocol for agreeing to retry dressing. If there is agreement from the user, then \RC{retry} is set to true and the transition back to \RC{Dressing} is enabled and taken. Otherwise, if no agreement has been reached and more than two minutes have elapsed since entering \RC{Abandoned}, the guard over the transition to \RC{Idle} becomes true. When that transition is taken there is a call to a software operation \RC{CallSupportDecision} that calls support depending on whether there is user assent for that.

The machine \RC{MonitoringService} also starts in an \RC{Idle} state, from where two outgoing transitions that can be triggered by events \RC{UserFallen} and \RC{CurtainOpenRqt}. If a user has fallen then the operation \RC{CallSupportDecision} is called, followed by the opening of curtains in the \RC{entry} action of the state \RC{OpenCurtains}, and then there is a transition back to \RC{Idle}. If there is a request to open the curtains via \RC{CurtainOpenRqt}, then there are two readings of measures \RC{userUnderDressed} and \RC{userDistressed} to determine if the user is under dressed and their level of distress. If the user is neither under dressed nor highly distressed, then the curtains are opened, and otherwise the request is refused as indicated by the output event \RC{refuseRequest}.

% %%%%%%%%%%%%%%%%%%%%%%%%%%%%%%%%%%%%%%%%%%%%%%%%%%%%%%%%%%%%%%
% %%%%%%%%%%%%%%%%%%%%%%%%%%%%%%%%%%%%%%%%%%%%%%%%%%%%%%%%%%%%%%
% \subsection{SLEEC rules for the case studies}
% %%%%%%%%%%%%%%%%%%%%%%%%%%%%%%%%%%%%%%%%%%%%%%%%%%%%%%%%%%%%%%
% %%%%%%%%%%%%%%%%%%%%%%%%%%%%%%%%%%%%%%%%%%%%%%%%%%%%%%%%%%%%%%

Robotic assistive dressing raises multiple SLEEC concerns. We show in Table~\ref{table:dressing-sleec} four rules defined by experts; their motivation, in terms of SLEEC principles that should be followed and why, is indicated in the table. 

\textsf{Rule1} is concerned with the time taken by a dressing episode; at most 2~minutes to complete, unless the room temperature is low, in which case it should be faster. \RC{Rule2} regulates the opening of the curtains, which should consider the privacy of the user, but also be sensitive to the distress that can be caused if a user request for the curtains to be opened is denied. If a user fall is detected then support must to called, but \RC{Rule3} requires that assent from the user is available for this. Finally, if the dressing is abandoned, there should be an attempt to retry, and, eventually the support must be called. 
Again, however, as required by \RC{Rule4}, the user's assent should be gained beforehand.

\begin{table*}[t]
    %\vspace*{-3mm}
\sffamily
\small
\setlength{\tabcolsep}{4pt}
    \centering
     \caption{SLEEC rules for the RAD system} \label{table:dressing-sleec}
     \vspace*{1.5mm}
    \begin{tabular}{p{1.1cm}p{10.5cm}p{1.8cm}p{3.7cm}}
    \toprule
    \textbf{Rule id} & \textbf{SLEEC Specification} & \textbf{SLEEC \newline principle}  & \textbf{Implication} \\ \midrule
   Rule1 & \begin{lstlisting}[language = SLEEC, label={Dressing:R1}, aboveskip=0pt,belowskip=0pt]
when DressingStarted then DressingComplete 
    within 2 minutes  
    unless roomTemperature < 19 then DressingComplete 
        within 90 seconds
    unless roomTemperature < 17 then DressingComplete 
        within 60 seconds
   \end{lstlisting}  & empathetic \newline ethical  & promotes and supports user\newline  well-being 
    \\ \midrule
   Rule2 & \begin{lstlisting}[language = SLEEC, label={Dressing:R2}, aboveskip=0pt,belowskip=0pt]
when CurtainOpenRqt then CurtainsOpened 
    within 60 seconds 
        unless userUnderDressed then RefuseRequest 
            within 30 seconds
        unless userDistressed > medium then CurtainsOpened 
            within 60 seconds \end{lstlisting}  & cultural \newline empathetic & respect for privacy and \newline cultural sensivity\\ \midrule % double check with the meaning 2 minutes does not make sense.
   Rule3 & \begin{lstlisting}[language = SLEEC, label={Dressing:R3}, aboveskip=0pt,belowskip=0pt] 
when UserFallen then SupportCalled
    unless not assentToSupportCalls \end{lstlisting} &  legal \newline ethical \newline social & respect for autonomy and\newline  preventing harm \\ 
   \midrule %confusing 
   Rule4 & \begin{lstlisting}[language = SLEEC, label={Dressing:R4}, aboveskip=0pt,belowskip=0pt] 
when DressingAbandoned then {RetryAgreed 
    within 2 minutes
        otherwise {SupportCalled
            unless not assentToSupportCalls}} \end{lstlisting} &  legal \newline ethical  & promoting user beneficence\newline  and respecting autonomy\\
\midrule
    \bottomrule
    \end{tabular}
\end{table*}

Only \textsf{Rule3} and \textsf{Rule4} have overlapping alphabets of events:~both refer to \RC{SupportCalled}.  So, using our SLEEC tool from Section~\ref{section:tool} we need to check them for conflict first.  As there is none, we need to check whether either of them is redundant.  They are not, so we can check next whether the design conforms to all rules. 

\begin{figure*}
    \centering    
    \includegraphics[scale=.75]{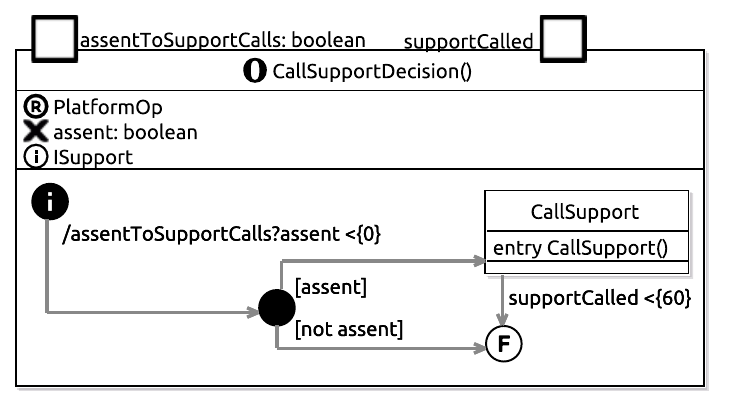}
    \caption{RoboChart model for \RC{CallSupportDecision}}
    \label{figure:callsupportdecision}
\end{figure*}

We get confirmation that the first three rules are satisfied, but \textsf{Rule4} is not.  The issue is related to the design of the operation \RC{CallSupportDecision()}, shown in Figure~\ref{figure:callsupportdecision}. In this version, the design engineer has followed an extra requirement to call the support in no more than 1~minute, if the user falls and there is consent. So, in \RC{CallSupportDecision()} there is a deadline on \RC{SupportCalled}, which flags the return from a call to an operation \RC{CallSupport()} of the platform.  With this deadline, we require that any actions involved in implementing \RC{CallSupport()}, such as establishing a phone connection and dialling, are completed within 1~minute.  Since \RC{CallSupport()} is a platform operation, it is asynchronous and does not block. 

The extra requirement is incompatible with the requirement to satisfy both \textsf{Rule3} and \textsf{Rule4}, because \textsf{Rule4} requires a delay of 2 minutes before calling support in case dressing is abandoned. There is no conflict between \textsf{Rule3} and \textsf{Rule4}, because \textsf{Rule3} does not impose a deadline on calling support. The extra requirement, however, creates a conflict. The counterexample, shown below, reveals the issue. Specifically, running the FDR model checker on the tock-CSP semantics generated by our SLEEC tool produces the trace: 
\begin{displaymath}
  UserFallen,DressingStarted,
  \\
  assentToSupportCalls.true, CallSupport,
  \\
  roomTemperature.-2, DressingAbandoned
    % DressingStarted, UserFallen, roomTemperature.-2,
    % \\
    % assentToSupportCalls.true, CallSupport
\end{displaymath}
This trace indicates that \RC{DressingService} has gone through \RC{DressingStarted}, got the measure -2 for the \RC{roomTemperature}, but finally \RC{DressingAbandoned} occurs. In \RC{MonitoringService}, \RC{UserFallen} has led to a call to \RC{CallSupportDecision()}, where  \RC{assentToSupportCalls} was found to be true, so a call to support is triggered, and then a deadline requires \RC{SupportCalled} to take place in 60 seconds. At this point, however, \RC{SupportCalled} is forbidden by \RC{Rule4} for two minutes so that a \RC{RetryAgreed} has a chance to occur. 

In this situation where the system design violates a SLEEC rule, we have to consult the SLEEC and requirements stakeholders. A few outcomes may be possible. A domain expert may agree that a one-minute deadline is too strict, and, in this case, the design may be changed. If \RC{DressingAbandoned} happens before \RC{SupportCalled}, RAD may consider whether the user is well enough to agree to a retry, in spite of having fallen.  In either case, \RC{SupportCalled} occurs after two minutes, either because the \RC{RetryAgreed} does not happen, or as a later response to \RC{UserFallen}.

If the solution agreed is to comply with the shorter time in the case of a fall, this can be captured by considering the RAD capabilities to call support in the case of a fall and to call support when dressing is not possible as distinct. In this case, we can have two different events representing two types of call to support. In fact, this may represent the fact that the information to be passed on to the support team in the different cases is different.  We may also have different support teams to deal with a fall, and with a difficulty to get dressed.